\documentclass[12pt]{article}
\def\id{\protect{{1 \kern-.28em {\rm l}}}}

\def\be{\begin{equation}}
\def\ee{\end{equation}}

\def\bea{\begin{eqnarray}}
\def\eea{\end{eqnarray}}

\def\p{{\partial}}

\makeatletter
\renewcommand\section{\@startsection {section}{1}{\z@}%
                                   {-3.5ex \@plus -1ex \@minus -.2ex}%
                                   {2.3ex \@plus.2ex}%
                                   {\normalfont\large\bfseries}}
\renewcommand\subsection{\@startsection{subsection}{2}{\z@}%
                                   {-3.25ex\@plus -1ex \@minus -.2ex}%
                                   {1.5ex \@plus .2ex}%
                                   {\normalfont\normalsize\bfseries}}

\makeatother

\def\eps{{\epsilon}}

\def\Tr{{\rm Tr}}

\def \foot {\footnote}
\def \bi{\bibitem}

\def \ha {{1 \over 2}}
\def \td {\tilde}
\def \ci{\cite}
\def \N {{\mathcal N}}

\def \d {\del}

\def \L {\Lambda}

\def\z{\zeta}

\def\b{\beta}

\def\p{\phi}

\def \del{\partial}

\def \aa {{\a'}}
\def\g{\gamma}

\def\z{\zeta}

\def\ov{\over}

\def\b{\beta}

\def\eps{\epsilon}

\def\foot{\footnote}

 \def \third { \textstyle {1\ov 3
}}
\def\det{\hbox{det}}
\def \ci {\cite}

\def \d {\partial}
\def \Tr {{\rm Tr}}
\def \P {\Phi}

\def \N {{\mathcal N}}

\def \td {\tilde}

\def \D {\Delta}
\def \N {{\mathcal N}}

\def \bi{\bibitem}
\def \la {\label}

\def\foot{\footnote}

\newcommand{\rf}[1]{(\ref{#1})}
\def \ov {\over}

\def\N{{\cal N}}

\def\cc{\circ}

\def \ha{{1\ov 2}}

\def \no {\nonumber}

\def \del {\partial}

\def \bi{\bibitem}
\def \la {\label}

\def\foot{\footnote}

 \def \p {\phi}

\def \ov {\over}

\def \varpi {{\rm w}}

\def \ep {\epsilon}

\def \DD {{\rm D}}

\def\Tr{{\rm Tr}}
\def \d {\partial}

\def\eps{{\epsilon}}

 \def \n {\nu}

\def \vp {\varphi}

\def \ed {\end{document}}

\def \te {\textstyle}

\def \ha {{{\textstyle{1 \ov2}}}}
\def \fo {{\textstyle{1 \ov4}}}

\def \D  {\Delta }

\def \ket  {\rangle}
\def \bra  {\langle}

\def \d {\delta}

 \def \eqref  {\rf}

\def \ha {{\textstyle {1 \ov 2}}}

\renewcommand{\theequation}{1.\arabic{equation}}
\def \bi {\bibitem}

\def \be {\bea}
\def \ee {\eea}

\def \dn {{\hat \d}}
  \def \d {\Delta}
 \def \DD {{\cal D}} \def \bP {{\bar P}}
\def \na {\nabla}

\def \ss {{\rm s}}\def \pe {\perp} \def \tri {{\te { 3 \ov 2}}}
\def \iffa {\iffalse}

  \def \De {\Delta}

\def \ads  {{(A)dS$_4$}}

 \def \kk {{\rm k}}

\def \emp {\emptyset}

\def \fiv  {{5\ov2}}
 \def \triv  {{3\ov2}}
 \def \hav  {{1\ov2}} \def \dva {\hav}
\def \aa  {{\rm a}}
\def \cc {{\rm c}}

\def \vp {\varphi}

\def \p {\phi} \def \L {\Lambda}
\def \D {{\rm D}}
\def \MM  {{\rm M}}

\def  \iffa {\iffalse}
\def \tri {\triv}
 \def \mm  {{\rm m}}

\def \ep {\epsilon}

\def \text { }

\def \pe {\perp}

\def \De {\dn}
\def \np {\newpage}

\begin{document}


\overfullrule=0pt
\parskip=2pt
\parindent=12pt
\headheight=0in \headsep=0in \topmargin=0in \oddsidemargin=0in

\vspace{ -3cm}
\thispagestyle{empty}
\vspace{-1cm}

\rightline{Imperial-TP-AT-2013-4}
 \def \G {\Gamma}

\begin{center}
\vspace{1cm}
{\Large\bf
On  partition function and Weyl  anomaly \\
\vspace{0.33cm}
   of  conformal higher spin  fields
}
\vspace{1.5cm}

 {
 A.A. Tseytlin\footnote{Also at Lebedev  Institute, Moscow. e-mail:
tseytlin@imperial.ac.uk }
}

\vskip 0.6cm

{
\em
\vskip 0.08cm
\vskip 0.08cm
Blackett Laboratory, Imperial College,
London SW7 2AZ, U.K.
 }
\vspace{.2cm}
\end{center}

\begin{abstract}
We  study 4-dimensional  higher-derivative  conformal higher spin (CHS) fields   generalising  Weyl graviton
and conformal gravitino. They
appear, in particular, as ``induced''  theories in the AdS/CFT context.
We consider their partition function on  curved Einstein-space   backgrounds
like  (A)dS or sphere    and Ricci-flat spaces.
Remarkably,  the  bosonic  (integer spin $s$)   CHS   partition function  
  appears to be
 given  by  a product  of partition functions of  the standard 2nd-derivative  ``partially massless''  spin $s$  fields,
 generalising the previously  known expression  for  the 1-loop Weyl graviton ($s=2$)  partition function.
 We compute the corresponding  spin $s$   Weyl anomaly  coefficients
 $\aa_s$   and $\cc_s$.  Our result  for $\aa_s$  reproduces    the  expression
   found   recently in  arXiv:1306.5242  by an indirect method   implied  by    AdS/CFT
    (which  relates   the   partition function of  a CHS  field on $S^4$  to a ratio  of  known  partition functions of
       massless higher  spin  field in AdS$_5$ with alternate boundary conditions).
 We also  obtain similar  results for the fermionic  CHS  fields.  In the half-integer $s$
 case  the   CHS partition function on (A)dS background
 is given by   the  product of  squares of  ``partially massless''    spin $s$  partition functions and one extra
 factor  corresponding  to  a special massive   conformally invariant  spin $s$   field.
 It was noticed  in arXiv:1306.5242
 that the sum of the bosonic $\aa_s$  coefficients over all $s$   is zero when  computed using the
 $\z$-function regularization,    and we  observe
   that the same property is true also in the fermionic case.

\end{abstract}

\newpage
\setcounter{equation}{0}
\setcounter{footnote}{0}

\section{Introduction}


Free Lagrangians   of massless  spin $\ha$  fermion  and   vector  spin 1    fields are  conformally invariant in $D=4$.
This  property is not shared by the standard  spin $\tri, 2$  and higher spin   massless fields.
In view of  potential    importance  of  the conformal invariance  condition
  one may  wonder if there  are alternative
higher-spin models  sharing  the conformal invariance  property with the  familiar $s=\ha,1$  theories.
Giving up manifest unitarity,  one may indeed view
higher-derivative Weyl graviton and conformal gravitino \ci{csg,bdw}  as   such  $s=2$ and $s=\tri$
 examples.  Conformal higher spin (CHS)  models  \ci{ftr}  are  their  $s> 2$ generalisations.
Regardless possible more  fundamental role of such   theories   \ci{ftr,frli}
they   naturally  appear  in the context of AdS/CFT  correspondence  \ci{lt,tse,seg,etc,mett,bek,gio}.\foot{In particular,
CHS fields are ``sources''  for higher  conserved conformal currents in the free limit of the boundary conformal
theory and  there are ``kinematical''    relations  between ordinary  massless higher  spins  in the bulk of   AdS$_{D+1} $   space and    conformal  higher spins     on the $D$-dimensional boundary   discussed, e.g., in
 \ci{re,bek2,vass} (for some other discussions  of conformal higher spins see also \ci{vasa,jm}).
  There are also
connections  of  conformal (super)gravity with   twistor string theory \ci{witt} and  with
scattering in dS  space \ci{mald,mai}.
}

CHS  theory   describes  pure spin $s$   field   by  a
{\it local}  action    with maximal  gauge  invariance.
Locality  requires  higher derivatives (unless one gives up minimal field content  requirement
and introduces auxiliary  fields):
free kinetic  operator is $ \D_s= \del^{2s} P_s$  where $P_s$   is pure spin $s$
(transverse traceless)  field projector.
When coupled to a metric, i.e. considered on a curved background,
the corresponding  action   should   be  given  by
  order  $2s$  Weyl and  reparametrization covariant  differential operator
   $\D_s(g)= \na^{2s} P_s + ...$  that
 generalizes  the $s=2$  4-derivative operator   appearing in the  linearization of the Weyl  gravity 
 action or  the   3-derivative  $s=tri$  conformal  gravitino operator.

 The corresponding   quantum 1-loop effective action  (given by  $\log \det\, \D_s(g) $)
 will not, in general,  be Weyl-invariant -- there will be a  non-trivial  Weyl  anomaly.
The well-known  low   spin $s= \ha, 1,  \tri,   2$   conformal field
 examples  (massless fermion, Maxwell vector,  Weyl  graviton and  conformal gravitino)
have non-zero  Weyl anomaly coefficients found   for
 $s=\ha,1$  in  \ci{dch} (see also \ci{cr1,cr2})     and  for $s=\tri, 2$  in \ci{ft0,ft1,ftr}.

The key  consistency   requirement  is  the  preservation of all
gauge symmetries, including the conformal symmetry,
 at the quantum level and thus  the cancellation of the Weyl anomaly \ci{ftr}.
This may be possible to achieve by  combining  several higher spin  fields together.
The computation of    Weyl anomaly  in conformal supergravities
\ci{ft1,ft2}   led to the conclusion that the only  known Weyl-invariant theory   with spins $\leq 2$ 
is $\N=4$ conformal supergravity  \ci{bdw}
coupled to $\N=4$    super Yang-Mills theory with  any   gauge group of rank 4.
An interesting   question  is  if there are  other special CHS models  involving spins $ \geq 2$    that  are
also  conformal  at the quantum level.\foot{One option
 may be  to   sum  over infinite number    of higher spin   modes
 (like in ordinary massless spin  theories of  Fradkin-Vasiliev type in AdS space),
 or one may try to explore possible existence of     anomaly-free  irreducible   models
  with    finite number
   of higher spins  like hypothetical $ N \geq  5$  conformal supergravities \ci{ftr}.}

To  find the  Weyl anomaly  of a  CHS field   in $D=4$
one would   need to  start   with the  corresponding  Weyl-covariant
   $2s$-derivative kinetic operator
in a background  metric and compute  the corresponding  heat kernel   coefficient
$b_4$   (often called also $a_2$).
One  immediate problem
is that   such   kinetic operator in  curved   metric  is  not known
explicitly so far  for $s >2$.\foot{Existence of conformal supergravity
implies that $s=2$ and $s=\tri$  kinetic operators  are consistent (have background gauge invariance)
provided the background metric satisfies  the Weyl gravity equations of motion
(e.g.,  is an Einstein-space metric).
Existence of    interacting conformal higher-spin theory was explored in \ci{seg,bek}.
 Consistent  cubic coupling of two  higher spins with a   conformal spin 2 field or the metric
 (e.g.,  via a  product of  two generalized Weyl tensors and  standard linearized Weyl tensor)
   implies that  higher spin
  gauge invariance should  hold provided  the metric satisfies the  linearized Weyl gravity equations of motion.
  In general, one   may  expect that CHS kinetic operator in background metric will
  have consistent  background   gauge invariance provided   the metric satisfies
   the non-linear Weyl gravity equations.
   We are grateful to K. Mkrtchyan and M. Vasiliev for  useful  remarks on this issue.
}
Moreover,  even if  such  kinetic operators   were  known, computing  their Weyl-anomaly    coefficients
would  be extremely  complicated  given the absence of general  higher-derivative  algorithms for $b_4$
 for $s>2$ (cf. \ci{ft0,ft1}).

This   suggests to take   a short-cut  route
originally    used  in   \ci{tss,ft3,ftr}
   as an efficient way to   reproduce  the  Weyl  anomaly  of conformal  graviton and conformal gravitino.\foot{See  \ci{pa} for a recent use of similar idea in $D=6$ context.}
First, to find the {\it two}  $D=4$ Weyl anomaly coefficients ($\aa$ and $\cc$) it is   sufficient  to consider
 just  two  {\it particular}  curved  Einstein-space
 backgrounds (e.g., conformally-flat  (A)dS$_4$ or $S^4$ one  and a Ricci-flat one).
 Second, in the case of Einstein-space background  one
 may expect   the covariant  higher-derivative  operator  to {\it factorize}
  into the  product of standard  2-nd order differential operators   whose anomaly can be computed using the standard algorithm.
  This   factorization was indeed  observed for the conformal graviton and conformal gravitino
  on the Ricci-flat  and  on conformally-flat   background  \ci{tss,ft3,dn1,dn2}.\foot{This factorization is not surprising
  given   that the Weyl gravity  Lagrangian $R_{mn}^2 - \third R^2$
  can  be written  in terms of the ``ordinary-derivative''   fields as
   $ 2 R_{mn} \p^{mn} +  R \p   -    \p_{mn}^2   + 3   \p^2$
where  $\p_{mn}$ and $\p$    are  the  auxiliary  traceless  tensor   and scalar  fields.
Similarly, the conformal gravitino  action
 can be written in terms of   three auxiliary  spin $\tri$ fields, symbolically, $\bar \vp D \vp  + \bar \psi D \psi   +
 \bar \chi ( \vp - D \psi) $ where $\vp$  will be the    gravitino field strength on-shell \ci{csg} (see also \ci{metr}).}

One may expect that   such  factorization  should  happen in general for  CHS  fields
on Einstein-space backgrounds. In flat space one may  replace (e.g., in bosonic case)  a higher-derivative CHS
Lagrangian  $\p_s \del^{2s} \p_s$ by  ``ordinary-derivative''  Lagrangian
  $\sum_{n,m}  \big(  c_{nm} \vp_n \del^2 \vp_m +  \mu_{mn} \vp_m \vp_n \big)   $
  for a  set of    fields  $\vp_n$  (some of which are of course ghost-like)
     such that
  integrating all   but one  of them  leads  back to the original  higher-derivative action.
  Such  gauge-covariant  ``ordinary-derivative''   formulations of
  both  bosonic and fermionic CHS  models in flat space were explicitly
   constructed in \ci{metcb,metcf}. Applying conformal  transformation (by redefining the fields
by powers of the conformal factor according to their  conformal weights)
one should then  get the
   corresponding action on a  conformally-flat   background.
   We shall assume that such actions
   should exist   also   on a  Ricci-flat   background.

Then  for    (A)dS or Ricci-flat    backgrounds   the higher-derivative
 CHS  kinetic  ($2s$ derivative bosonic or squared fermionic)
operator   should { factorize}   into product  of  standard  2nd-derivative  operators, so that the CHS
  Weyl anomaly can be found as a    sum  of   Weyl anomalies of  an effective set of  ``ordinary-derivative''  fields.

\iffa
The    method  we shall use  was  originally    suggested in   \ci{tss,ft3,ftr}
   as an efficient way to   reproduce  the  Weyl  anomaly  of conformal  graviton and conformal gravitino.
   It is based on   the following    key observations:

   (i) since  in $D=4$ the Weyl  anomaly  is determined by the {\it two} coefficients $(\aa,\cc)$
to find them    it is sufficient to  consider  two  {\it particular}  curved  Einstein-space
 backgrounds (conformally-flat one  and Ricci-flat one);

(ii) for such  special     curved backgrounds   the higher-derivative   CHS  kinetic ($2s$ derivative bosonic or squared fermionic)
operator  {\it factorizes}   into product  of  standard  2nd-derivative  operators, so that the CHS Weyl anomaly
 can be found as a
sum  of   Weyl anomalies of  an effective set of  ``ordinary''  fields.
 \fi


   In more detail,  in 4 dimensions   the  non-trivial part of the Weyl  anomaly
is determined by \ci{desa} (we  omit  the $\na^2 R$ term)
\bea \la{1}
&&b_4 =   \b_1  R^*R^*   + \b_2 W =   - \aa\,  R^*R^*   + \cc\, C^2        \ ,\\
\la{1a}
 &&    \aa = - \b_1 +  \ha  \b_2   \ , \ \ \ \ \ \ \    \cc= \ha  \b_2   \  , \ \ \ \ \ \ \ \   \b_1 = \cc- \aa \ , \ \ \  \b_2 = 2   \cc  \  .
\eea
Here $  R^*R^*$    is  $ 32 \pi^2$ times  the Euler number integrand    and $C^2$  is the  square of the Weyl tensor,
\be \la{2}
C^2 = R^*R^* +  2 W \ , \ \ \ \ \ \ \      W= R^2_{mn} - {{\te {1 \ov 3}}} R^2  \ .
\ee
 The integral of  $b_4$   is related to  the coefficient  of the UV logarithmic
  divergence  of the  corresponding curved-space  partition function\foot{Here $\det\,  \D$  is  assumed  to include ghost contributions and
  $L\to \infty $ is UV cutoff. Overall sign is   changed in the   case of fermions.}
\be
\ln Z = - \ha   \ln \det\, \D = \ha \big( \ha L^4   B_0   + L^2  B_2  +    B_4 \ln L^2  \big) + {\rm finite}  \ , \ \ \ \
B_p = {\te{1 \ov (4 \pi)^2}} \int d^4 x \sqrt g \ b_p  \la{4}. \ee
Here $b_0 = \nu$  is the   number of effective degrees of freedom and $b_2=0$
 for conformally covariant operators.
 The known values of  the anomaly coefficients for low conformal spin cases $s=1, 2$  and $s={\te { 1\ov 2},{3\ov 2}}$ are:\foot{For standard real conformally-coupled scalar
 $\b_1  = {   1 \ov 180} ,\   \b_2  = {  1 \ov 60} $,
 while for 4-derivative real conformal scalar \ci{ft1} \  $\b_1  = {   1 \ov 90} ,\   \b_2  = -{  2 \ov 15} $.
 Let us
 recall also that  for  standard
$\N=4$ conformal supergravity $\b_1=0, \ \b_2=-2$ so that coupling it to exactly four   $\N=4$ SYM  multiplets
(that have $\b_1=0,\ \b_2= \ha$)   leads to  anomaly-free theory \ci{ftr}. The  cancellation is possible  due to the fact that $\aa$ and $\cc$
of conformal   gravitino have  negative  sign.}
 \bea
 &&\te s=1: \ \ \    \b_1  =- {   13 \ov 180} \ ,\ \   \   \b_2  = {  1 \ov 5} \  , \ \ \ \ \ \  \ \ \  \aa=  {   31 \ov 180} \ ,\ \  \ \  \ \cc= { 1 \ov 10} \ , \la{44}\\
 &&\te s=2: \ \ \    \b_1  = {   137 \ov 60} \ ,\ \   \  \ \  \b_2  = {  199 \ov 15} \  , \ \ \ \ \ \  \    \aa=  {   87 \ov 20} \ ,\ \ \ \ \ \ \ \cc= { 199 \ov 30} \ , \la{45}\\
 &&\te s=\ha: \ \ \    \b_1  = {  7 \ov 720} \ ,\ \   \ \ \   \b_2  = {  1 \ov 20} \  , \ \ \  \ \ \ \ \      \aa=  {   11 \ov 720} \ ,\ \  \ \  \ \cc= { 1 \ov 40} \ , \la{46}\\
 &&\te s={\te {3\ov 2}} : \ \ \    \b_1  =- {   173 \ov 180} \ ,\ \   \   \b_2  =- {  149 \ov 30} \  , \ \ \  \ \  \aa=  -{   137 \ov 90} \ ,\ \   \ \cc=- { 149 \ov 60} \ . \la{47}
 \eea
 To find  the   two   anomaly coefficients it should  thus  be   sufficient  to compute   the  logarithmically
 divergent part   of the partition
function on    two   particular  curved backgrounds that have different values of $R^*R^*$ and $C^2$.  
Obvious  examples  are  provided by  the   Einstein spaces
\be R_{mn}=  \Lambda  g_{mn}  \ ,  \la{ee} \ee
  which  solve the Weyl gravity equations of motion
and, as was already mentioned above,
 should thus allow for a consistent higher-spin coupling.\foot{To  have gauge independence  of  partition function of Weyl gravity in curved background   
one is to expand near solutions of the Weyl theory equations  of motion. 
All   solutions of  \rf{ee}
     are  solutions of these equations.
Same  applies  to conformal gravitino:  the existence of   conformal supergravity \ci{csg}  implies   
consistent coupling of  conformal  gravitino to Weyl gravity    and thus
its partition function  is   gauge-independent  provided the background metric satisfies
the Weyl gravity equations of motion.  It is natural to assume that  all CHS
 fields  can   be  consistently coupled to   Weyl gravity, and then their
 kinetic   operator should   have background gauge invariance  provided  the  metric satisfies Weyl equations, and, in particular, \rf{ee}.}
It is   sufficient to consider  the two special   cases:
a conformally-flat  Einstein   space (de Sitter $\ep >0$ or Anti de Sitter   $\ep <0$)\foot{We shall mostly consider $D=4$ case
with Minkowski signature (for some notation and useful relations see Appendix )  but all results will have straightforward
continuation to the Euclidean signature case: in particular,  (A)dS$_4$   can be replaced by
 $S^4$  background.}
\be
 C_{mnkl} =0 \ , \ \ \ \ \ \ \    R= 4 \L= 12 \ep  \ , \ \ \
 \ \ \ \ \ \    b_4 =- \aa\, R^*R^* =- { 24  }\aa\,  \ep^2     ,  \la{5} \ee
and  a Ricci-flat space
\be   
R_{mn} = 0 \ ,  \ \ \ \ \ \ \    \ \ \ \ \ \ \   b_4=   \b_1 R^*R^*  = (\cc-\aa) \,   C^2 \ .  \la{6} \ee
Computing $b_4$ for   (A)dS$_4$   background  will determine $\aa$ and
then computing   $b_4$   for a Ricci-flat background   will   allow  also  to find    $\cc$.

As discussed above, the second
  key simplification is that for such special   backgrounds the CHS kinetic operators should factorize  into products
of ``ordinary-derivative'' operators whose Weyl anomaly  can be  computed using the standard algorithm \ci{alg}.
In (A)dS$_4$   background    \rf{5}  this factorization was observed   for  the  Weyl  graviton and conformal gravitino
 in \ci{tss,ft3}  and nearly simultaneously   in \ci{dn1,dn2}.
For example, the Weyl graviton operator factorizes into the product of the  usual ``massless''  transverse traceless graviton operator  and a  massive transverse traceless spin 2  operator.
The latter  was  noticed  in \ci{dn1}   to  have     special  conformal invariance property  and
was    called   ``partially  massless''  spin 2   field.
 ``Partially  massless''  (PM)  higher spin  fields   in
(A)dS$_D$   exist also for $s > 2$
 \ci{hig,dw1,dw2,z1,dnw,dw3,vasil,mef} (see also \ci{all,deswj,gri})  and, as we shall suggest,  are directly
    related to the factorization of the CHS operators  on  (A)dS$_4$   background
    for all $s$.

\iffa
The  factorization of  CHS  kinetic operator   in  conformally-flat background should indeed be expected.
First, the Weyl invariance of the CHS action implies  that  it can be transformed (by redefining the fields
by powers of the conformal factor according to their  conformal weights) into flat space. The  flat-space action
for a CHS field  can be  rewritten  in an ordinary-derivative (2nd for bosons and 1st for fermions) form
by introducing extra auxiliary fields (some of which are of course ghost-like) \ci{metcb,metcf}.
This ``ordinary-derivative'' approach  should have generalization to generic  curved background (or at least Einstein-space background)
 as  was  already  apparent  in \ci{csg}  for conformal graviton and gravitino.
 \fi

Starting with the  ``ordinary-derivative''  formulation  \ci{metcb,metcf}, applying a conformal transformation
  and  solving for  all of  the auxiliary fields
one should  end up  with the factorized form of the CHS kinetic operator  in the (A)dS$_4$  background.
As we shall  suggest  below,   the $s$ of the     2nd-derivative  operator  factors   that form
 the  $2s$-derivative    bosonic   spin $s$   CHS   kinetic  operator   can be   identified  precisely   with
 the $s$ species  of  the  PM   spin $s$   fields  with the  mass  parameters
  found in \ci{dw1,dw2,z1}.\foot{A relation between a general $s >2$ conformal higher spin  field and partially massless  
  fields in   (A)dS  was  anticipated  by   E. Skvortsov and M. Vasiliev  (unpublished)  and also  conjectured in 
  \ci{jm}.}
 As we shall see, a  similar   relation  appears to
  exists also  between  the fermionic CHS operator  and the  fermionic PM   fields \ci{dw2,mef}    in  (A)dS$_4$  plus
   an additional massive spin $s$  field.

As was  noted above,  kinetic operators for the  Weyl  graviton and the conformal gravitino  factorize  also in a Ricci-flat background
  \ci{ft3,ftr}: here the factorization is even simpler  -- into the  relevant  powers (second and third) of   the
  standard  covariant massless  graviton  and massless   gravitino operators. 
  Given that CHS kinetic operator  is defined on traceless and transverse symmetric tensors
  it is natural to expect that   the same   factorization  pattern should   apply also for all CHS  cases   with  $s >2$.

Assuming   these factorization relations   and including the relevant ghost  determinant factors
 one  is then able to express   the  CHS  partition  function on  a  conformally-flat or
  Ricci-flat   background in terms of  products of
 powers of determinants of ordinary 2nd-derivative  operators and thus to
  compute  the corresponding  values of the
 Weyl   anomaly  coefficients   using the standard algorithm for $b_4$    for  $\Delta_2 =-\na^2 + X$  type   operators.
 This is the   strategy that we  will  implement  below.

 Our  results  that generalize the low-spin expressions in \rf{44}-\rf{47}    can be summarized as follows.
 In the bosonic integer spin $s=1,2,3,...$ case the Weyl anomaly coefficients in  \rf{1}  for
  a  CHS  field read
  \bea
 &&  \aa^{(b)}_s=  { \te{1 \ov 720}}\, \nu_s \, ( 3\n_s  + 14 \nu^2_s)       \ ,\ \ \ \ \ \ \  \ \ \ \ \ \ \ \ \ \ \  \nu_s = \nu^{(b)}_s=s(s+1)    \  ,  \la{19} \\
  &&  \cc^{(b)}_s=  { \te{1 \ov 720}} \,\nu_s\, (  4 - 42  \n_s   + 29 \n_s^2 )         \ , \la{20}
  \eea
  while in the fermionic  case  of  $s=\ha, \tri, {\te{5\ov 2}}, ...$  we found
  \bea
 &&  \aa^{(f)}_s=  {\te{ 1 \ov  2880}}\, \nu_s \, ( 12 + 45\n_s  + 14 \n^2_s)       \ ,\ \ \ \ \ \ \  \ \ \ \ \ \ \  \nu_s =\nu^{(f)}_s=   -2 (s+\ha )^2  \  ,  \la{19f} \\
  &&  \cc^{(f)}_s=  { \te {1 \ov 2880} }\,\nu_s\, (  118  +135 \n_s   + 29 \n_s^2 )         \ .\la{20f}
  \eea
Here  $\nu_s$ is the number of dynamical degrees of freedom of the corresponding   CHS field
(with fermionic modes  counted with negative sign, i.e. $\nu_1=2, \ \nu_{{1\ov 2}} = -2$, etc.).

  Remarkably, the  bosonic $\aa_s$   coefficient   in \rf{19} that we will   find  below directly in $D=4$
    matches  the expression
  found  recently in  \ci{gio}  by a completely  different  indirect method  based on AdS$_5$/CFT$_4$ duality.\foot{Note that our  normalization   of $\aa_s$ is   different from the one  used in \ci{gio}    by  factor of 4.}
  Ref.\ci{gio}  started with  the  expected
   relation between  the  spin $s$  CHS  partition function on S$^D$   and the  ratio  of  the standard
  massless  higher spin $s$    partition functions  on AdS$_{D+1}$   \ci{gab,lal}
   with alternate boundary conditions  which is
   implied   by the consideration of  the RG flow  induced  by the
    ``double-trace''  deformation   \ci{rast,guk,har,dorn} of the boundary CFT  by the  square of the
   corresponding  spin $s$ conformal current.
   The conformal anomaly $\aa$-coefficient
   was  then  extracted from the singular
   $\ln L $  term   (cf. \rf{4}) in  the  predicted  CHS   partition function on $S^4$.\foot{The same  
    idea was  previously used  in \ci{diaz}  to reproduce  the $\aa$-coefficient in Weyl anomaly 
 of  the  conformal  4-derivative  scalar  operator \ci{ft1,ftr,pan} and of its higher-dimensional generalizations (see Appendix B).}

     It was observed in \ci{gio} that if one  sums $\aa_s$ in \rf{19} over all spins $s=1,2,...$ and  computes
     the (formally power-divergent) result  using the $\z$-function     regularization
       the final expression vanishes,
     suggesting that  a  theory  containing each   bosonic CHS  field just once may be  quantum-consistent.\foot{Such theory would
     be originating as an induced one starting with a large  $N$  free complex scalar CFT  with all
      $s>0$  currents being gauged.
     One may then expect the   induced CHS  theory to be consistent \ci{gio}, i.e.  at least do not have
    logarithmic  UV  divergence  in partition function  on
     a  sphere or  IR  divergence in the corresponding   ratio  of AdS partition functions with alternate boundary conditions,
     implying the vainsing of the  sum of the $\aa_s$   coefficients.}
         Interestingly,  the fermionic  CHS  coefficient $\aa_s$  in \rf{19f}   also has the  same property, i.e.\foot{Instead of  $\z$-function regularization  one may, in fact, use    any consistent analytic regularization (see discussion in section 5  below).
     Use of   such regularization
     in  computing  infinite sums over spins appears to be natural and necessary in  the  AdS/CFT  context, see
     \ci{gio,gk}.}
  \be
     \sum_{s=1}^\infty \aa^{(b)}_s\Big|_{\rm reg}=0 \ , \ \ \ \ \ \ \ \ \ \ \ \
     \sum_{s=\dva}^\infty \aa^{(f)}_s\Big|_{\rm reg}=0 \ . \la{fif} \ee
    At the  same time, similar  sums   of the $\cc_s$   coefficients  in \rf{20},\rf{20f}
    do not  appear to vanish (and thus same applies  also  the sums of the corresponding $\b_{1,s}=\cc_s-\aa_s$
    coefficients  in \rf{1}). That may be suggesting that  either  some subtlety
    was overlooked  in our computation of $\b_{1,s}$\foot{The  idea  of computation  of  $\b_{1,s}$-coefficient   using the
   assumption of  factorization  of $s > 2$ CHS operators  on Ricci-flat background came up  in our discussions with  S. Giombi  who  independently  found  \rf{20}.}
    or that  to achieve the vanishing  of the full Weyl anomaly
   one is to consider a particular   combination of the  (bosonic and fermionic)   CHS fields.

This  paper is organized as follows.
 We shall start  in section 2
 with  a   review  of CHS  fields in flat  space extending the discussion in \ci{ftr}
 and clarifying the structure of the corresponding free partition functions.

In section  3 we shall discuss the bosonic CHS  fields in curved space.
We shall  first  consider (in section  3.1)  the case  of a Ricci-flat background  and suggest a natural   factorized representation
for the quantum partition function for a spin $s$  CHS field  that generalizes both the
known conformal graviton expression \ci{ft1,tss,ft3} and the   flat space expression.
The  corresponding   $\b_{1}$   Weyl anomaly coefficient is then computed  as in \ci{alg,cr2}.

In section 3.2 we shall turn to the case of a  CHS   field  on  conformally-flat (A)dS$_4$    background
and will argue  that  the corresponding   spin $s$ partition function  can be  represented  as a the product of
$s$   factors  which are quantum partition functions of  the   standard 2nd-derivative  massless  and $s-1$
 ``partially massless''  spin $s$   fields   in (A)dS.
 These PM partition functions  correspond to quantization of Lagrangians in \ci{z1,mef}
 were  not given   previously  in the literature.
  This leads to the $\aa_s$ expression in \rf{19}, and, combined  with
 $\b_{1,s}$ found  in section 3.1, to the $\cc_s$ in \rf{20}.

The analysis  of the fermionic CHS  case   in section 4 is similar: we start with Ricci-flat case  and determine $\b_{1,s}$
(section  4.1),
and then turn  to the  conformally-flat   case. The partition function on (A)dS   background (section  4.2)
is again  expressed in terms of  the  massless  spin $s$ partition function and the product of ``partially massless''
ones  but here  there  is  also one extra factor which corresponds to a massive  spin $s$ mode without any residual gauge invariance. The resulting expression  is direct  generalization of the conformal gravitino partition function  in
\ci{tss,ft3,ftr}  and also of the   flat-space fermionic CHS  partition function. The corresponding $\aa_s$ coefficient
is given by \rf{19f}  and together with $\b_{1,s}$ from section 4.1 it leads to $\cc_s$ in \rf{20f}.

In section 5   we shall  make some concluding remarks
discussing in particular    the finiteness property of the sums  over spins  in  \rf{fif}
and the one-loop relation  \ci{gio} between  ratio of  massless  higher-spin determinants in AdS$_{D+1}$ 
and CHS partition function on $S^D$. 

Appendix   contains some definitions and useful relations.
In Appendix B we review    factorization  of conformal  higher derivative  scalar  operators
on Einstein space background into  product of  2nd-derirvative scalar Laplacians    which is analogous to the one discussed in section 3 for CHS operators.

\renewcommand{\theequation}{2.\arabic{equation}}
 \setcounter{equation}{0}
 \section{Flat space background}


 Conformal higher  spin theories   have    free kinetic Lagrangians  with  maximal degree of   gauge invariance
  and irreducibility consistent with locality, i.e. describe pure spin states   even  off shell  \ci{ftr}\foot{Note that here
 the scalar $s=0$  case   is that of   a trivial non-propagating field   with
 no dynamical degrees of freedom. There are of course 2- and 4-derivative  conformally-covariant
 scalar operators   but they are not natural members of  the CHS  family we are going to discuss.}
  \be
  L_s = \P_s\,  \D_s\,    \P_s \ ,  \    \ \ \ \ \  \qquad \D_s= P_s \,   \del^{2s}  \  . \la{7} \ee
 Here  we assume  that space-time dimension is $D=4$   and 
   for integer $s=1,2,...$   we have  bosonic field $\P_s=\hat \p_s=(\hat \p_{m_1...m_s})$ which 
 is real totally symmetric  tensor  of rank $s$,
 and $ \del^{2s}  = (\del^2)^s$.
 For half-integer  spin $s=\ss  + \ha$ ($\ss=0,1,2,...$)
 the  corresponding field is fermionic   $\P_s=\hat \psi_s=(\hat \psi_{m_1...m_\ss})$ which is
 Majorana  totally symmetric spinor-tensor   of rank $\ss=s- \ha $,
 and $ \del^{2s} = (\del^{2})^\ss   \gamma^m \del_m $.
 $P_s=(P^{m_1 .. m_s}_{n_1 ...n_s} )$ is totally symmetric, traceless ($\gamma$-traceless   in  half-integer spin  case)
 and transverse projector in each  of the two sets of indices.\foot{For comparison,
 to  get a   Lagrangian of  an ordinary higher spin field one is to  start with
  an operator  $\DD_s =\bar  P_s \del^2 $  for bosons   and    $\DD_s = \bar P_s \gamma^m  \del_m $  for fermions
  and choose  $\bP_s = P_s +  a_1  P_{s-1}  + ..$   as   particular  combinations of lower-spin projectors
 such that the resulting Lagrangian is local. For example,  in the Maxwell,  Einstein  and  the standard gravitino cases one has
 $\bP_1 =  P_1, \ \bP_2 = P_2 - 2 P_0, \ \bP_{\tri} = P_{\tri} - 2 P_{\hav} $.}
 
To find a formal generalization of  \rf{7} to   any  dimension $D$ (see, e.g.,  \ci{seg})  one is  to  
 replace   the kinetic operator $P_s \del^{2s}$  
   by 
   \be  \D_s = P_s \,   \del^{2s + D-4} \ , \la{ged} \ee
    i.e.  to shift   the power of the Laplacian 
 $s\to s + \ha (D-4)$.  The corresponding field $\Phi_s$   has dimension $2-s$  for all $D$.
  For  even $D > 4$   the $s=0$ scalar field will   have a higher-derivative 
 conformal kinetic operator $ (\del^2) ^{ D-4 \ov 2} $ (see Appendix B).

 The structure of \rf{7}  implies the presence of both differential (analogs of Maxwell or reparametrization)   gauge invariances
 with parameters $\hat \xi_{s-1}$  and algebraic (analogs of Weyl   or conformal supersymmetry) gauge invariances
 with parameters $\hat \eta_{s-2}$;  symbolically,
 $\delta \hat  \phi_s= \del\hat  \xi_{s-1}  + g_2 \hat \eta_{s-2}$  ($g_2$ stands for the metric factor).
 The presence of  higher-derivative $\del^{2s}$   factor
  in the  kinetic term  ensures  the locality of these actions.

 In what   follows we shall always    consider  only symmetric  {\it traceless}    tensors $\p_s$
 in the case of bosons  and
 symmetric  {\it $\g$-traceless}  tensors $\psi_s$ in the case of  fermions.
 Then  the algebraic part of the gauge group will be automatically taken into account
 and the remaining differential  gauge freedom  will be parametrized  by symmetric traceless  tensors $\xi_{s-1}$.
 The number of components of  a  totally symmetric traceless rank-$s$ tensor in $D$  dimensions  is
 \be
  N_s\equiv  N(\p_s) = {\te {\pmatrix{s+D-1\cr s}- \pmatrix{s+D-3\cr s-2}}} \  , \ \ \ \ \ \ \ \ \ \
  N_s\Big|_{D=4}   =  (s+1)^2  \ . \la{101} \eea

 \subsection{Bosons}

Let us   start with  bosonic CHS  fields   having  integer $s=1,2,3,...$. In what follows we shall  mostly  be interested in
the $D=4$ case  but  will quote some  relations for general $D$.
 The number of   off-shell  degrees of freedom  (i.e. the  number of components  minus  dimension of gauge group)
 of a bosonic CHS field of spin $s$  is given
  by
 \be \la{ne}
 N_{s\pe} =  N_s- N_{s-1} = 2 s+1   
   \ , \ee
   where $N_{s\pe}\equiv N(\p_{s\pe})$  is the number of components of transverse  ($\del \cdot \p_{s\pe}=0$) traceless rank $s$  tensor field.
  The number of dynamical  (or on-shell) degrees of freedom $\nu_s$  can be defined
  in terms of the corresponding free
 partition function 
 \be\la{8}
  Z_s =(\det \, \D'_s )^{-1/2} =  ( Z_0)^{\n_s} \ , \ \ \ \ \ \ \ \  \ Z_0 = (\det\, \del^2)^{-1/2} \ ,   \ee
 where  prime   indicates proper gauge fixing  and $Z_0$ is  a standard
 real scalar field  partition function.
 Since the action \rf{7} depends only on the transverse part of $\p_s$,   changing the variables $\p_s \to (\p_{s\pe}, \xi_{s-1})$
 as
 $\p_s =\p_{s\pe} + \del \xi_{s-1}$  with Jacobian $(\det \Delta_{s-1})^{1/2} $,
 where $\Delta_k=-\del^2$ is   defined on totally symmetric traceless  rank $k$ tensors,
 and dividing over the volume  of gauge group (i.e. omitting  spurious integral over $\xi_{s-1}$)  we find
 for the CHS partition function
 \bea \la{92}
 &&Z_s
 =      \Big[  { \det  \Delta_{s-1}   \ov ( \det \Delta_{s \pe} )^{s}         } \Big]^{1/2}
 \ . \ee
 Here $\Delta_{k\pe}=-  \del^2$  is  defined on symmetric traceless {\it transverse}  rank $k$  tensors.
 Using that
 \be  \la{98}
   \det  \Delta_{k} = \det \Delta_{k\pe } \  \det  \Delta_{k-1} \ , \ee
    we  can rewrite  \rf{92}  in two alternative forms:  in terms of   unprojected operators $\Delta_k$
     \bea \la{9}
 &&Z_s
 =      \Big[  { (\det  \Delta_{s-1} )^{s+1}   \ov ( \det \Delta_s )^{s}         } \Big]^{1/2}
 \ , \ee
or in terms of  transverse-projected   operators
     \be   Z_s =\prod_{k=0}^{s-1} Z_{s,k} \ , \ \ \ \ \ \ \ \ \
 Z_{s,k}\equiv    \Big[  { \det  \Delta_{k  \pe }  \ov \det \Delta_{s\, \pe} }\Big]^{1/2}  \ .
 \la{91}
 \ee
 Here the  indices of  $Z_{s,k}$   indicate  ranks of tensors on  which the two 2nd-order operators (in the denominator
 and numerator)  are defined.
 In particular, \be
 Z_{s,s-1}=   \Big[  { \det  \Delta_{(s-1)  \pe }  \ov \det \Delta_{s\, \pe}} \Big]^{1/2}=
  \Big[  { (\det  \Delta_{s-1} )^2  \ov \det \Delta_{s} \    \det \Delta_{s-2}  } \Big]^{1/2}    \la{93} \ee
 is the partition function of a  standard massless    higher spin   field.

 In general, $Z_{s,k}$  in \rf{91}    can be  formally  interpreted as a partition function of a spin $s$
 field  with  gauge invariance  $\delta \p_s=  \del^{s-k} \xi_k$ involving higher derivatives
 but lower-rank parameter tensor (becoming  the standard   $\delta \p_s=  \del \xi_{s-1}$  for $k=s-1$
  corresponding to \rf{93}).  The number of the associated   dynamical degrees of freedom   is then
  \be
  \nu_{s,k} = N_{s\pe} -  N_{k\pe} = 2 (s-k) \ , 
  \la{9111}   \ee
  where $ N_{s\pe}$  was  given in \rf{ne}.
 Thus $\nu_{s,k}$     ranges from $\nu_{s,0}=2s$    to   $\nu_{s,s-1} =2$   in the  massless   case.
 As follows from \rf{9} or \rf{91}, the  number of dynamical degrees of freedom
  of a  bosonic CHS field  is  then \ci{ftr}
   \be  \nu_s =  s N_s - (s+1) N_{s-1}   =  \sum_{k=0}^{s-1} \nu_{s,k}=  s (s+1) \ . 
   \la{10}
    \ee
 For example,  $\nu_1=2$ for a vector, $\nu_2=6$  for a  Weyl  graviton \ci{ft0,ft1}, etc.

 As we shall see below, the  representation \rf{91} of the flat-space
 partition function   has  a natural generalization to the case when  the CHS field
 is propagating in an Einstein-space  background, in particular,  (A)dS$_{4}$  one,
 when  $Z_{s,k}$  become  partition functions of {\it ``partially massless''}   fields.
  The representation \rf{9}   will  also have a  direct  generalization to a Ricci-flat background.

The expressions \rf{92},\rf{9},\rf{91}   have the following generalization to even $D >4$: 
\bea \la{922}
 Z_s
 &&=      \Big[  { \det  \Delta_{s-1}   \ov ( \det \Delta_{s \pe} )^{s + {1 \ov 2}  (D-4) }         } \Big]^{1/2}
 =      \Big[  { (\det  \Delta_{s-1} )^{s+1 +  {1 \ov 2} (D-4)} 
   \ov ( \det \Delta_s )^{s+  {1 \ov 2} (D-4) }         } \Big]^{1/2}\no \\
&&  =\prod_{k=0}^{s-1}  \Big[  { \det  \Delta_{k  \pe }  \ov \det \Delta_{s\, \pe} }\Big]^{1/2}  \
 \Big(  {  1    \ov  [\det \Delta_{s\, \pe} ]^{1/2}       } \Big)^{  {1 \ov 2} (D-4) }     \ . 
 \la{911}
 \ee

 \subsection{Fermions }

Next, let us   consider the case of  fermionic    CHS fields  with  half-integer  $s= \ss +\ha= \hav,\tri,\fiv...$.
The number of components of   Majorana  $\g$-traceless  (and thus also
traceless)  spinor-tensor $\psi_{s}=(\psi_{m_1 ...m_\ss})$ in $D=4$
is  \be
  N_\ss\equiv N(\psi_s) =  4\Big[ {\te {\pmatrix{\ss+ 3 \cr \ss}- \pmatrix{\ss+2\cr \ss-1}}} \Big] =  2( \ss+1) (\ss+2)= 2 (s + \ha  )   (s + {\textstyle {3\ov 2}} )   \ ,
  \la{1012} \eea
and thus  the number of off-shell degrees of freedom  is (cf. \rf{ne}) \ci{ftr} 
 \be \la{nef}
 N_{\ss\pe} \equiv N(\psi_{s\pe} )=  N_{\ss} - N_{\ss-1} =   4 (\ss + 1) =   2 (2s+1)
   \ . \ee
Squaring  the  fermionic CHS kinetic operator  $\del^{2 s} P_s$  and writing the corresponding  partition function in terms
of 2nd-order  Laplacians  we find the following analog of the bosonic  expression \rf{92}
  \bea \la{10f}
 &&Z_s
 =      \Big[ { ( \det  \Delta_{s-1})^2   \ov ( \det \Delta_{s\pe}  )^{2s  }         } \Big]^{-1/4}
 =      \Big[{  \det  \Delta_{s-1}  \ov ( \det \Delta_{s\pe}  )^{\ss } }\Big]^{-2/4}   \Big[ {1 \over  \det \Delta_{s\pe}       } \Big]^{-1/4}     \ ,
 \eea
 where  $\Delta_s=-\del^2$  is  defined on totally symmetric  $\g$-traceless   spinor-tensors  $ \psi_{s}$
 and we used  that  as in the    bosonic case   the
 Jacobian of transformation  from $\psi_s$ to $\psi_{s\pe}$ (i.e. $\psi_s= \psi_{s\pe} + \del \xi_{s-1}$)  produces  the  factor
 $ (\det  \Delta_{s-1})^{-1/2}$ (with  extra -1  power).
 Depending on  interpretation,  the  structure of the  expression
 \rf{10f}    is formally different from  that of \rf{92}  in  either  having $ \det  \Delta_{s-1} $   in the numerator  being squared
  or in having  an extra  factor of $\det \Delta_{s\pe} $.

 The equivalent forms of \rf{10f} are  (cf. \rf{9} and \rf{91})
  \bea \la{9f}
 Z_s
 =      \Big[  { (\det  \Delta_{s-1} )^{s+1}   \ov ( \det \Delta_s )^{s}         } \Big]^{- 1/4}
 \ , \ee
or
     \be   Z_s =\prod_{k=1/2  }^{s-1} (Z_{s,k})^2 \   Z_{s,\emptyset} \ , \ \ \ \ \ \ \ \ \
 Z_{s,k}\equiv    \Big[  { \det  \Delta_{k  \pe }  \ov  \det \Delta_{s\, \pe} }\Big]^{- 1/4}  \ ,   \ \ \ \ \ \ \
  Z_{s,\emptyset}\equiv    \Big[ { 1 \ov \det \Delta_{s\, \pe} }\Big]^{- 1/4} \ .
 \la{11f}
 \ee
Like in the bosonic case in \rf{93},  here $Z_{s,s-1}$   is
 the partition function of  the standard (1-st derivative)
 massless  fermionic spin $s$ field,   while  $Z_{s,k}$  may be  interpreted as  corresponding to a spin $s$ field
 with smaller  gauge group  but with $s-k$  derivatives in the gauge transformation.
  The corresponding number of fermionic
 d.o.f. is then
 \be
 \nu_{s,k}  = - \ha (  N_{\ss\pe}  - N_{\kk\pe}) = - 2 (s-k) \ , \
  \ \ \ \ \ \ \ \ \   \nu_{s,\emptyset}= - \ha   N_{\ss\pe}  = - (2 s +1)
     \ , \la{11ff} \ee
     where  $k=\kk + \ha, \ \kk = 0,1, ...$.
 A peculiarity of the fermionic  CHS case is the
  second power of $Z_{s,k}$ in \rf{11f} and the presence of  an extra
 ``purely-massive'' mode  (with no remaining gauge invariance)   represented by  the $Z_{s,\emptyset}$ factor.
 As in the bosonic  case, the representations  \rf{11f} will have  natural  generalizations  to
  curved Einstein-space backgrounds.

 The  fermionic CHS  field
    number of dynamical degrees of freedom following from  either  \rf{9f} or \rf{11ff}   is thus   (cf. \rf{10})
  \be
    \nu_s =  -   s N_s  +  (s+1) N_{s-1}     =   2\sum^{s-1} _{k= 1/2} \nu_{s,k}  +   \nu_{s,\emptyset}   = - 2 (s+ \ha)^2= - 2 ( \ss + 1)^2
     \ . \la{11} \ee
    For example,  this gives  the standard    values  for spin 1/2  field and  spin 3/2 (conformal gravitino) fields:
       $\nu_{\hav} = -2$   and $\nu_{\triv} = - 8$  \ci{ftr}.
       Explicitly,  according to \rf{11f} the  partition function of  conformal gravitino $Z_\triv$   is
        a product of   2 partition functions  of ordinary massless gravitino  and one  ``purely massive''   gravitino,
        implying  $-(2\times 2 +4)=-8$  for the d.o.f.  count.
  For   $s= \fiv$    CHS field we get   2 massless  spin  $\fiv$   factors $Z_{\fiv,\triv}$, 2  ``partially massless''
  spin $\fiv$  factors $Z_{\fiv,\hav }$  and one ``purely  massive''   factor $Z_{\fiv,\emptyset}$, etc.

 For an $\N=1$   CHS  supermultiplet containing   spins $\{s, s+\ha , s+1\}$  (i.e. $ \{\ha,1\},\,  \{1, \tri, 2\}$, etc.)
   one  then finds from \rf{ne},\rf{10},\rf{11}    the expected   result:   $\sum_{\{s\}} n_s=0, \ \ \sum_{\{s\}} \nu_s=0$ \ci{ftr}.

\renewcommand{\theequation}{3.\arabic{equation}}
 \setcounter{equation}{0}
\section{Bosonic conformal higher spins in curved background}


Here we shall    consider the expressions for   the   integer $s$  CHS partition functions in
Einstein-space (Ricci-flat and conformally-flat backgrounds)
starting  first  with  the known cases  of low spins $s=1$ (Maxwell vector)  and $s=2$ (Weyl graviton) 
and then   suggesting natural generalizations to $s>2$.

 \subsection{Ricci-flat background}

The Maxwell   vector partition function   in a curved background has familiar form
 \be
  Z_1 =   \Big[{\det \d_0 \ov \det \d_{1\pe}}\Big]^{1/2}=  \Big[ {(\det \Delta_0 )^2 \ov   \det  \Delta_1       } \Big]^{1/2}   \ , \ \ \ \ \ \ \ \
  \Delta_0  = - \na^2 \ , \ \ \ \   ( \Delta_1)_{mn} = - ( \na^2)_{mn}  + R_{mn}   \ . \la{3.1}
 \ee
 Considering a conformal spin $s>1$  field  in   an external metric (i.e. coupled to $s=2$ conformal  field)
 one,  in general,  should get  a  complicated  $2s$-derivative  reparametrization and Weyl
 covariant  differential operator  with coefficients
 depending on the background  curvature. However, this  operator may simplify  for specific backgrounds,
 reducing to a product of lower-dimensional   operators.

 This indeed happens  for the 4-derivative Weyl graviton operator in  Einstein-space backgrounds.
One  finds  \ci{tss,ft3,ftr}   that  in  the   Ricci-flat  background     the  Weyl
 graviton 4-th order operator factorizes, becoming the square of the
traceless Einstein  graviton operator. Then     the   Weyl gravity 1-loop partition function
takes formally  the same   form as its flat-space  counterpart in  \rf{9}. It
 can be  expressed also  as a product  of the  familiar one-loop partition functions  of the two
  Einstein gravitons   \ci{haw}
  and  one Maxwell   vector  in  the $R_{mn}=0$ background:
 \bea \la{12}
 &&Z_2 =  \Big[{ \det \d_0 \ov (\det \d_{2\pe})^2 }\Big]^{1/2} = \Big[{ (\det \d_1)^3 \ov (\det \d_2)^2 }\Big]^{1/2} =   (Z_{2,1})^2 \,  Z_1  \  ,
 \\  && Z_{2,1}= \Big[{\det \d_{1\pe} \ov \det \d_{2\pe}}\Big]^{1/2}
  = \Big[ {(\det \d_1)^2  \ov \det \d_2\  \det \d_0}\Big]^{1/2}
    . \la{12a}
\eea
Here
 $\d_0  $ and $   (\d_1)_{mn}$ are as in \rf{3.1},  
   the operator
  $    (\d_2)_{mn,kl} = - \nabla^2 _{mn,kl} - 2 R_{mknl} $
is assumed to be defined on traceless symmetric  2-tensors  and $\d_{k\pe}$ are defined on transverse symmetric traceless  tensors.

Note that $Z_2$ in  \rf{12}  and  $Z_1$ in  \rf{3.1},\rf{12a} have  the same structure as
 the flat-space partition function \rf{9}    but   here  with  covariant differential operators $\d_s$.
 It is then   natural to expect that  for any conformal higher spin field in a Ricci-flat background   the   kinetic operator should  be factorizing into $s$  factors of the  ``massless''  spin $s$ 2nd-order operators. Then
     the partition function should be given by
\be
Z_s =   \Big[{ (\det \d_{s-1} )^{s+1}  \ov (\det \d_s)^s }\Big]^{1/2} \ ,   \la{12b}\ee
where $\d_k$  are    covariant  2nd-order differential  operators  defined on traceless rank $k$ tensors
corresponding to standard massless
spin $k$ fields.
This  appears to be  the simplest ``minimal''-coupling assumption extended to the CHS  fields.

In general, one may  define  (following, e.g.,  \ci{cr1,cr2})  an operator   acting on a
field in an  irreducible  $SO(1,3)$  representation
$(A,B)$ of dimension $ N_{(A,B) }=(2A+1)(2B+1) $  as ($A,B$ are positive half-integers)
\be \la{jo}
\Delta_{(A,B)} = - \nabla^2(V) +  X \ , \ \ \ \ \ \ \ \ \  \ \ \  X= - R^{ab}_{\ \ mn} \Sigma_{ab}  \Sigma^{mn} \ .
\ee
Here $V_m= \omega^{ab}_m \Sigma_{ab}$, \  $\omega^{ab}_m$ is the standard spin connection
 and $\Sigma_{ab}$ are the  corresponding generators  of $SO(1,3)$.
 In  the  present case of symmetric traceless rank $s$ tensor  fields we have
 \be
 (A,B)= ({\te {s\ov 2}}, {\te {s\ov 2}}), \ \ \ \ \ \  N_{(A,B) }\equiv  (2A+1) ( 2 B+1) =  N_s = (s+1)^2 \ , \ \ \ \ \ \
\Delta_{(A,B)}  = \Delta_s   \ , \la{boo} \ee
where $\Delta_s $   is defined on symmetric traceless  rank $s$  tensors
  and  corresponds to  standard ``harmonic-gauge''   massless  higher spin operators
  on a  curved  background, generalizing the $s=2$ Lichnerowitz operator. 
   Explicitly \ci{por1,cort},  for the symmetric traceless tensor representation the Lorentz generators are 
  $(\Sigma_{mn})^{a_1....a_s}_{\ \ b_1....b_s} = s \delta^{(a_1}_{ [m} \eta_{n](b_1} \delta^{a_2}_{b_2} ...\delta^{a_s)}_{b_s)}$
  so that 
  \be  (X \phi_s)^{a_1...a_s} = - s(s-1)  R_{m\ \ n \ }^{ \ (a_1 \ a_2 \ } \phi^{a_3 ...a_s mn }    + s   R_{m}^{\ (a_1}  \phi^{a_2 ...a_s) m }\ . 
  \ee
  As follows from \rf{12b},    one can then express  the value of the    $\b_{1}$   coefficient  in \rf{1}
 for a  CHS  field as a   combination of $\b_{1}$  coefficients for the operators $\d_s$
 by a relation similar  to the one for  the number of degrees of freedom \rf{10}
  \be   \b_{1,s}   =  s\,  \b_1[\d_s]  - (s+1)\,  \b_1[\d_{s-1}]   \   ,    \la{ww}
  \ee
  where $\b_1 ( \d_{k}) $    is the $\b_1$ coefficient  in the expression \rf{1} for $b_4(\d_k) $.
  The  latter    can be computed using the  standard  algorithm \ci{alg}   for  the 2-nd order operators
   $\Delta = - \nabla^2(V) + X$   defined  in curved space on fields $\P^i$ with  $\nabla_m$   containing
  the matrix   connection $(V^i_{\ j})_m$:\foot{In the present cases $V_m $ will be  expressed in terms of the
  spin  connection
   so that  $\Tr F^2_{mn}(V)$   will    give contraction of two curvatures.}
   \be \la{algo}
   b_4[\Delta]
   = \te { 1 \ov 180} \Tr \Big[  15 F^2_{mn}(V) + 90 X^2  - 30 R X  - 30 \nabla^2 X  +
     {\bf 1}  ( R^*R^* + 3 R^2_{mn}  + { \te{3 \ov 2}} R^2 + 6 \nabla^2 R ) \Big]
   \ee
Following \ci{cr1,cr2} one finds  in the Ricci flat case  ($\Tr  {\bf 1}  = N_{(A,B)} $) 
 \be \la{21}
\b_1[\Delta_{(A,B)}]  = \te{1 \ov 180}  N_{(A,B)}
 \Big( 1  + A(A+1) [ 6A(A+1) -7]   +  B(B+1)[ 6 B(B+1) -7 ) ] \Big)  \ ,
\ee
 so that in the present  case of \rf{boo}
 \be
 \b_1[\Delta_{s}] =\te {1 \ov 720}  N_{s } ( 21 - 20 N_s   + 3 N_s^2)  \ . \la{211}\ee
 Finally,  for   the  bosonic CHS  field  the $\b_1$  coefficient \rf{ww}  is thus given by
 \be
   \b^{(b)}_{1,s}   =     \te {1 \ov 720} \n_s ( 4 - 45 \n_s + 15 \n_s^2) \
            ,     \  \ \ \ \ \ \ \ \ \ \ \  \n_s = s(s+1)   \  .  \la{www}
  \ee
 This  expression  agrees with the  known  values for   $s=1$ and $s=2$
 in \rf{44},\rf{45}  (and it vanishes as it should  for non-dynamical $s=0$  case).

  It is interesting to note that  while on general grounds (cf. \rf{algo})
  the   Weyl anomaly  coefficients   in   \rf{1}  for a  CHS field  of  spin $s$  should
be   6-th order polynomials  in $s$,  $\b_1$   is actually  a  cubic  polynomial  in  the number of
dynamical degrees of freedom $\nu_s=s(s+1)$.   The same  will apply also to the
expression for the  second Weyl anomaly  coefficient $\b_2$  discussed below.

 \subsection{Conformally-flat background}

Next, let us determine $\aa$  in \rf{1}  (and thus $\b_2$)
   by considering  a   constant  curvature (A)dS$_4$   background \rf{5}.
The  Maxwell   vector  partition function \rf{3.1}  in this case  may be written as
\be  Z_1 =
\Big[ {     \det \dn_0 (0)      \ov   \det \dn_{1\perp} ( 3)  }  \Big]^{1/2}=
 \Big[ {     (\det \dn_0 (0) )^2     \ov   \det \dn_{1} ( 3)  }  \Big]^{1/2} \la{14a} \ .
 \ee
Here   and in what follows  the operator
\be\la{33}   \dn_{s} (M^2 ) \equiv   - \nabla^2_s  +  \MM^2 \ , \ \ \ \ \ \ \  \ \ \ \ \ \ \   \MM^2 \equiv M^2\ep  \ee
will be   defined on symmetric traceless tensors   and
$\dn_{s\perp} (M^2 )$ will stand for $   \dn_{s} (M^2 ) $
 defined on {\it transverse}  symmetric traceless tensors.
 The  parameter  $\ep= \pm {1 \ov r^2} $ is   equal to 1
  for  unit-radius dS space and  -1
  for  unit-radius  AdS space.\foot{Note also that the
   notation $M^2$ does not mean that
 this dimensionless parameter is  always  positive.}

As in the Ricci-flat case, one finds that the Weyl graviton kinetic operator again
 factorizes \ci{tss,ft3,dn1,dn2}:\foot{Note,  in particular, that
$\d_2 \p_{mn} =   - \nabla^2  \p_{mn}   -2 R_{mknl}\p_{kl}  \to  \dn_2 (2)  \p_{mn} = (- \nabla^2_2  + 2 \ep )\p_{mn}$.}
  \be  C^2 =  \ha \phi_2\,  \dn_{2\perp} (2 )\,   \dn_{2\perp} (4 )\,  \phi_2   + O(\p_2^3)  \ . \la{wey}  \ee
As a result,  the 1-loop partition function of the Weyl  theory  can be written as
\ci{tss,ft3,ftr}\foot{For some general relations between
 $\dn_{s} $ and  $\dn_{s\perp}$ operators see Appendix.}
\be \la{13}
&&Z_{2 }  
 = Z_{2,1} Z_{2,0} =  \Big[ {  \det \dn_{1\perp} ( -3 )  \ov   \det \dn_{2\perp} (  2  ) }  \Big]^{1/2} \
  \Big[ {   \det \dn_0 (-   { \te{ 4 } })      \ov  \det  \dn_{2\perp} ( { \te{ 4 } })    }  \Big]^{1/2}
\ . \la{133}
\ee
For $\L=3\ep\to 0$  the mass terms  in \rf{33}  disappear
and   this reduces to the flat-space expression in \rf{91}.

In contrast to the flat and Ricci-flat case in \rf{12}
here the two  standard graviton operator  factors  are not  the same: the degeneracy is lifted by the curvature.
 The first  factor
  is the usual
 (A)dS$_4$  ``massless''    graviton  contribution
   equal to  the    1-loop partition function  of the Einstein gravity  with cosmological term   \ci{gp,cr3}
\be \la{14}
Z_{2,1}=   \Big[ {  \det \dn_{1\perp} ( - 3)      \ov   \det \dn_{2\perp} ( { \te{ 2}} )    }  \Big]^{1/2}= \Big[
 {  (\det \dn_1 ( - 3)  )^2       \ov     \det \dn_{2} ( { \te{ 2 }} )\   \det \dn_0 ( - 6 ) }  \Big]^{1/2}  \ .
\ee
The second   factor
\be \la{142}
Z_{2,0} =
  \Big[ {   \det \dn_0 (-   { \te{ 4 } })      \ov  \det  \dn_{2\perp} ( { \te{ 4 } })    }  \Big]^{1/2}
= \Big[{  \det \dn_1 ( - 1) \   \det \dn_0 (-   { \te{ 4} }) \ov
\det  \dn_{2} ( { \te{ 4 } })      \ }  \Big]^{1/2}  \ \ee
   corresponds to  the   ``partially massless''  spin 2
 field found
  in \ci{dn1,dn2,dw3} to have a special conformal covariance property
  (allowing to transform its equation of motion to the  massless
 flat space $\del^2$ form  and thus  ensuring its ``null-cone'' propagation).
 This
 field has on-shell 2nd-derivative
 gauge invariance  with a scalar  parameter,
   $\delta \p_{mn} = ( \nabla_m \nabla_n  +  4 \ep   g_{mn} ) \xi$ \ci{dw1,dw2}.
 This field  can be described by a local Lagrangian \ci{z1}  involving  the  standard   2-tensor and vector fields
 with gauge invariance $\delta \p_{mn}  = \nabla_{(m} \xi_{n)}   +  \mu    g_{mn} \eta, \ \
 \delta A_{m}  = \nabla_m \eta   +   \mu      \xi_m, \ \mu^2 = -4 \ep $ (so that  it effectively describes
 same number  of dynamical
 d.o.f. $4=2+2$  as a  massless   spin 2  plus  a  massless   spin 1  system). Quantization of this
 system leads  to the partition function \rf{142}.\foot{This  can be   made  more apparent
  by   rewriting  the  partition function
 \rf{142}   as
 \be \no
 Z_{2,0} =
  \Big[ {     \det \dn_{1\pe}  ( - 1) \       \ov  \det  \dn_{2\perp} ( { \te{ 4 } })    }  \Big]^{1/2}\
    \Big[ {\det \dn_0 (-   { \te{ 4 } })  \ov    \det \dn_{1\pe}  ( - 1) } \Big]^{1/2} \ .
\ee
}

Partially massless  (PM)  fields  in \ads\  that admit local gauge-invariant description
after introduction of some extra lower-spin   modes  exist     for all  $s >2$. 
For given  value   of $s$   there is  one massless  and $s-1$   PM fields    which in the
general case of (A)dS$_D$
with  curvature given  in \rf{a1}     are described by the following  kinetic operators \ci{dw2,z1}
\be
&& \dn_{s} (M^2_{s,k})  = - \na^2_s    +  M^2_{s,k}\ep   \ , \ \ \ \ \ \
  \ \ \ \ \ \ \   k=0, 1, ..., s-1  \la{pm} \ ,
   \\
    &&   M^2_{s,k}=  s - (k-1) ( k + D-2)  \ , \ \ \ \ \ \ \ \
    M^2_{s,k}\Big|_{D=4} =  2 +s  - k - k^2 \ .  \la{maa} \ee
Here  $k=s-1$   corresponds to the massless  field   in  (A)dS$_D$  
\ci{fro,met}
with the mass parameter\foot{Separating the massless
spin $s$ contribution to $M^2$ one may   write the mass  parameters of the PM fields as
$ \ \ \ \ M^2_{s,k}= m^2_{s0}  + \mu_{s,k}^2 \ , \ \ \
\mu_{s,k}^2 = (s- k-1) ( s+ k + D-4) .$
%
Discussion of massless fields in AdS$_D$ in the frame-like, metric-like, and BRST approaches may be found in the respective references \ci{mebm}.
%
}
\be
m^2_{s0} \equiv
M^2_{s,s-1}=  s  - (s-2) (s + D-3)   \ , \ \ \ \ \ \ \ \ \ \ \          M^2_{s,s-1}\Big|_{D=4} =  2 + 2 s -s^2 \ .      \la{ppm} \ee
Interestingly,  the   ``transpose''  of \rf{maa}
\be   M^2_{k,s}=  k - (s-1) ( s + D-2)   \ , \ \ \ \ \ \ \ \
    M^2_{k,s}\Big|_{D=4} =    2 +k  - s - s^2   \   \la{kaa} \ee
    gives
\be
M^2_{s-1,s}=  - (s-1) (s + D-3)   \  , \ \ \ \ \   \ \ \ \          M^2_{s-1,s}\Big|_{D=4} =  1-s^2  \ ,   \la{mm} \ee
 which is  exactly  the mass parameter of the ``ghost'' factor  in the partition function of the massless
spin $s$ field    in  (A)dS$_D$   \ci{gab,lal}
\be
Z_{s,s-1} = \Big[ {\det  \dn_{s-1\, \pe} (M^2_{s-1,s}) \ov\det  \dn_{s \pe} (M^2_{s,s-1}) } \Big]^{1/2} \ . \la{mas} \ee
This  of course agrees with  the  $s=1,2$   expressions   \rf{14a} and \rf{14}  in  $D=4$.\foot{Note also
    that the
``maximal-depth''   PM field  with $k=0$ (in $D=4$)      plays a somewhat
 special role being conformally-invariant \ci{dw3}
 and  having the
highest-derivative  (order $s$)  gauge invariance  with a scalar gauge parameter.
A discussion of  this field and some   hints of its connection to CHS fields    appear
 in \ci{gri}.
}

Our key  observation  is that   CHS   kinetic operator in  conformally-flat background should factorize
into  precisely $s$   factors  of  ``partially massless''    kinetic operators \rf{pm}, i.e.
\be \p_s {\rm D}_s \p_s = \p_s  \Big[ \prod_{k=0}^{s-1} \dn_{s\perp} (M^2_{s,k})    \Big]   \p_s \  , \la{pass}\ee
thus generalizing  the  familiar
Maxwell and Weyl theory \rf{wey}   cases.
One   possible derivation of this  relation  may  start from
 the  ``ordinary-derivative''   formulation of the CHS theory in flat space  \ci{metcb}, then  explicitly
transforming to conformally-flat  metric and  finally solving for  all  auxiliary
 fields.

To obtain  the  quantum
 CHS partition function in a  conformally-flat background  it remains then to find the corresponding
  ``ghost'' factor.
As in  the low-spin  and massless   spin examples  \ci{gp,cr3,all,fto,tss,ft1,ft4,camp,gab,lal}  it   is  found  using
  the Jacobian of transformation from the traceless   field
$\p_s$ to its   transverse component  $\p_{s\pe}$  and  other lower-spin transverse fields (cf.  Appendix
for some examples).
One is also to divide over the
volume of  the gauge transformation group with {\it unconstrained}  traceless  parameters.
The final  expression for the   CHS  partition function in (A)dS$_4$   background
then  takes the following remarkably simple form
which is  a generalization of the flat-space expression in \rf{91} 
\be
Z_{s} =\prod_{k=0}^{s-1}  Z_{s, k} \ , \ \ \ \ \ \ \ \ \
Z_{s,k}=  {  \Big[ {\det  \dn_{k \pe} (M^2_{k,s}) \ov \det \dn_{s \pe} (M^2_{s,k}) } \Big]^{1/2}} \ . \la{fi} \ee
Here the $k=s-1$   factor is  precisely  the massless  spin $s$  partition function \rf{mas}
and other factors correspond to the PM fields.

As  follows from the structure of the flat-space partition function in \rf{911}, to find a generalization of \rf{fi}  to  general even dimension $D$  one needs to  multiply    \rf{fi} (now with $D$-dependent massess 
\rf{maa},\rf{kaa})  by  
extra $\ha (D-4)$  ``purely-massive''  (no residual 
gauge invariance)   factors.\foot{The presence of such extra massive    degrees of freedom 
in the $D\not=4$ case is implied also by the structure of the ``ordinary-derivative'' 
formulation of \ci{metcb}  (we are grateful to R. Metsaev for pointing this out).}
These  have the following general 
form:
\be \la{exx}  Z_{\rm extra}= \prod^{  D-4\ov 2}_{i=1} \Big[ {1\ov  \det \dn_{s \pe} (m^2_{s,i})} \Big]^{-1/2}\ , 
\qquad \ \ \ \ \  \dn_{s \pe} (m^2_{s,i}) =  - \na^2_s    +  m^2_{s,i}\ep\ , 
\ee  where 
 the  mass  coefficients  $m^2_{s,i}$ remain to be determined. 
 They are  actually  known in the special case of $s=0$  when  the  CHS operator becomes 
 the conformal  scalar operator $\Delta_{(2r)}$   with $r= \ha (D-4)$  (see Appendix B):
  from \rf{b4}   we have\foot{For  $s=0$ 
 the product in \rf{fi} is to be set to 1 as there $k \geq 0$  and $Z_{0,0}=1$.}
  \be 
  m^2_{0,i} =  - ( i- \ha D ) ( i + \ha D  -1)   \ , \ \ \ \ \ \ \ \ \      i= 1, ..., \ha (D -4) \ .  \la{ex1} \ee
 Comparing this with ``partially-massless''  mass formula \rf{maa} 
 with $s=0$ we observe  that  it  coincides with \rf{ex1}   if we set 
 $k = i- \ha D + 1$, i.e. $i=1, ..., \ha(D-4)$ correspond to $k=-\ha (D-4), ..., -1$. 
 Then 
  a natural 
 conjecture is that in general  one should have 
 \be 
  m^2_{s,i} = s  - ( i- \ha D ) ( i + \ha D  -1)   \ , \ \ \ \ \ \ \ \ \      i= 1, ..., \ha (D -4) \ ,   \la{ex11} \ee
 i.e. the  massive factors   in \rf{exx}   may be interpreted as the 
 ``partially-massless''    contributions  with massess  $M^2_{s,k}$  in \rf{maa}  extended to {\it negative}
  values of $k =  - \ha (D-4),  ..., -1$ (and without ``ghost''  factors $ \det  \dn_{k \pe} (M^2_{k,s})$ 
   in \rf{fi}).\foot{There  appears to be a group-theoretic  argument leading 
   to  this relation (E. Skvortsov and M. Vasiliev, private communication).}


The   partition function  \rf{fi}  can be   written also   in terms of unconstrained operators
as in \rf{14a},\rf{133}  using the following relation (valid for any $k=1,2,...$)
\be \la{ddi}
 \det \dn_{k\perp}(M^2)   = { \det \dn_{k}(M^2 ) \ov   \det \dn_{k-1}  ( M^2- \delta_k) } \ , \ \ \  \ \ \ \ \ \
  \delta_k= 2 k + D-3  \ , \ \ \ \ \ \   \delta_k\Big|_{D=4} = 2 k +1  \ .
\ee
For example,  for the  massless factor \rf{mas} we then get
\be
Z_{s,s-1} = \Big[ { (\det  \dn_{s-1} (M^2_{s-1,s}))^2   \ov\det  \dn_{s} (M^2_{s,s-1}) \  \det  \dn_{s-2}
(M^2_{s+2,s+1})  }      \Big]^{1/2} \  ,              \la{mast} \ee
where we used  that
\be   M^2_{s,s-1} - \delta_s = M^2_{s-1,s} \ ,  \ \ \ \ \ \ \ \ \    M^2_{s-1,s}   - \delta_{s-1} = M^2_{s+2,s+1} =
2- s^2 - s(D-2)  \ . \la{yy} \ee
In particular, for $s=1$ and $s=2$  and $D=4$    the  expression \rf{fi}  agrees  with \rf{14a} and \rf{13}.
Also, for $s=3$ in $D=4$   we find  from \rf{maa}  that  $M^2_{3,k}= 5 -k -k^2 , \ \ M^2_{k,3} = k-10$ 
where $k=2,1,0$   and thus\foot{Note that here and in \rf{13}  we write the factors in the opposite order to \rf{fi} so that the massless
spin factor appears first.}
 \be \la{134}
&& Z_{3 }= Z_{3,2} Z_{3,1} Z_{3,0} =
\Big[ {   \det \dn_{2\perp} ( -8 ) \ov  \det \dn_{3\perp} ( { \te{ -1}} )\ }\Big]^{1/2} \
\Big[ {   \det \dn_{1\perp} ( -9 ) \ov  \det  \dn_{3\perp} ( { \te{ 3 } })}\Big]^{1/2} \
\Big[ {  \det \dn_0 (-   { \te{ 10 } })          \ov
   \det  \dn_{3\perp} ( { \te{ 5 } })  }  \Big]^{1/2}
 \\  &&
=  \Big[ {    (\det \dn_{2} ( -8 ))^2 \ov    \det \dn_{3} ( { \te{ -1}} )\  \det \dn_{1} ( { \te{ -13}} )\     }        \Big]^{1/2}
      \    \Big[ {    \det \dn_{2} ( -4 ) \ \det \dn_{1} ( -9 ) \ov  \det  \dn_{3} ( { \te{ 3 } }) \   \det  \dn_{0} ( { \te{ -12 } }) }
 \Big]^{1/2}       \    \Big[ {  \det \dn_{2} ( -2 ) \ \det \dn_0 (-   { \te{ 10 } })     \ov
  \det  \dn_{3} ( { \te{ 5 } })  }  \Big]^{1/2}.\no
 \ee
In the second line we used \rf{ddi}.  Here  the first  factor  is
 the massless spin 3 partition function  (cf. \rf{mast}).
Note that in the limit $\ep\to 0$  all spin 0 and spin 1 factors cancel and we
recover the flat-space expression in \rf{9}.

It is  now straightforward to apply the $b_4$ algorithm \rf{algo}  to each of the 2nd order operator  
in the  CHS partition function \rf{fi}
 in  conformally-flat  $D=4$
space
 to compute the corresponding $\aa$-coefficient according to \rf{5}.
Let us  first consider  a generic   unconstrained operator \rf{33} defined on symmetric traceless tensors
  with  an arbitrary   dimensionless  ``mass''  constant   $M^2$  and
 with $\na_m=\na_m(V)$   with  connection $V_m$  corresponding to an $SO(1,3)$
 representation $(A,B)$  (in particular, to the one in  \rf{boo}).
In the conformally flat Einstein-space   background case  we get from \rf{algo}    (here $R=12\ep$)
\be  b_4[ \dn_s ( M^2 )] =\te  { 1 \ov 180} \Tr \Big[ 15 F^2_{mn} (V)  + {\bf 1}  (90 M^4   - 360 M^2        +  348  )\ep^2    \Big] \ , \la{h1}
\ee
where  \ci{cr2}
 \be
 \la{f1}  \Tr  F^2_{mn}(V)  = - 4 N_{(A,B)}  [ A(A+1) + B (B+1)]  \ep^2 \ .  \ee
 Using \rf{boo} we  find \
 $ \Tr  F^2_{mn}(V)   = - 2 N_s  s(s+2) \ep^2 $, \ \ $N_s= (s+1)^2$.
Then we  may determine  the  contribution to the
$\aa$-coefficient  corresponding  to the operator  \rf{33} according to \rf{5}
\be  \aa[ \dn_s (M^2)] =   \te { 1 \ov 144}    N_s   \Big(  N_s     - 3  M^4    + 12 M^2
 -  {\te {63\ov 5} }    \Big) \ . \la{aa2}
\ee
Using \rf{ddi}   we may then find also the $\aa$-coefficient
 corresponding to the  transverse operator as (cf. \rf{ne})\foot{The  use of the  $b_4$  for the  unprojected operator
 means that we are effectively computing the anomaly of the  partition function expressed in terms of  unprojected operators
 like in the second line  of \rf{134}, thus avoiding a subtlety with  zero-mode  contributions if one computes 
 the anomaly using  $\zeta(0)$  for the  projected operators
    (cf. \ci{fto,tss,ft3}).}
  overall factor here
 \be &&  \aa[ \dn_{s\pe} (M^2)] =  \aa[ \dn_{s} (M^2)] -  \aa[ \dn_{s-1} (M^2 - 2s-1)]
\no \\ && \ \ \ \ \ \ \ \  \ \ \ \ \ \ \ =
\te   \te { 1 \ov 720}   (2s+1)    \Big[  30 s^3 + 85 s^2  + 10 s - 58     - 30  (s^2-2) M^2     - 15  M^4     \Big] \ . \la{aa22}
\ee
It  is now straightforward  to compute the resulting  total value of the $\aa$-coefficient
corresponding to the CHS partition function \rf{fi}   with the mass parameters given in \rf{ppm}.

Computing  the finite  sum over $k$ as implied  by  the representation \rf{fi}  we  end up with a
simple expression  for the bosonic CHS  anomaly   coefficient $\aa_s$
\be
\aa^{(b)}_s&=&\sum_{k=0}^{s-1}   \Big(   \aa[ \dn_{s\pe} (  2 +s  - k - k^2     )]     -        \aa[ \dn_{k\pe} ( 2 + k  - s - s^2    )] \Big)
 \no\\ &=& 
          \te{ 1 \ov 720} \n_s^2 ( 3 + 14 \n_s)  \ , \ \ \ \ \ \ \ \ \ \ \ \ \ \ \ \    \nu_s= s(s+1) \ .  \la{nen}
\ee
Like the $\b_1$ coefficient  found earlier in \rf{www},
 it depends   on $s$ only  through  the corresponding
 number of dynamical  degrees of freedom  $\nu_s$ in \rf{10}.
As already mentioned in the Introduction,   \rf{nen}
 matches the expression  for $\aa^{(b)}_s$ found   in \ci{gio}  by an indirect method.

For comparison, the contribution of just massless   spin $s$ part of  \rf{mas} (with 2 degrees of freedom)  is\foot{The same
 expression was
recently used  in \ci{gk},  leading to the conclusion
 that the logarithmic  divergence   in  the 1-loop partition function of (type-A/B) Vasiliev's
higher-spin theories  in AdS$_4$ vanishes  assuming  one uses the
 $\zeta$-function  to  define  the sum over all  integer spins $s$.}
\be
\aa^{(b)}_{s,s-1} =\te {1\ov 360}  (2 - 15 s^2 + 75 s^4) \ . \la{cha}
\ee
It of course agrees   with \rf{nen}  for $s=1$ when
 \ $\aa_1= \aa_{1,0} = {31\ov 180} $.

Combining the results   for $\b^{(b)}_{1,s}$ \rf{www}   and $\aa^{(b)}_s$ \rf{nen}   we conclude from  \rf{1a} that
\be
\cc^{(b)}_s = \ha \b^{(b)}_{2,s}= \b^{(b)}_{1,s} + \aa^{(b)}_s
 =  \te{ 1 \ov 720} \n_s (4- 42 \n_s  +   29  \n^2_s)  \ . \la{nyn}
\ee

  \renewcommand{\theequation}{4.\arabic{equation}}
 \setcounter{equation}{0}


   \section{Fermionic  conformal higher spins  in curved background}

   Let us   now consider  the   fermionic CHS   fields  with half-integer  spin $s= \ss +\ha$   ($\ss= 0,1,2, ...$)
   described  by  symmetric  $\g$-traceless   spinor-tensors  $\psi_{m_1...m_\ss}$.
   We shall follow the same  strategy as in the previous  section, first discussing  the Ricci-flat   and then
   the conformally-flat  backgrounds.

 \subsection{Ricci-flat  background}

 The simplest  examples  of the fermionic CHS   fields   are    the   Majorana spinor\foot{We represent the partition
 functions in terms of squared fermionic operators.}
 \be
 Z_{{1\ov 2}} = \Big[ { 1 \ov \det \d_{{1\ov 2}}} \Big]^{-1/4} \ , \ \ \ \ \ \ \ \
 \d_{{1\ov 2}}  = - \nabla^2_{1\ov 2}   +\te  { 1 \ov 4} R \to  - \nabla^2_{{1\ov 2}} \ , \la{121} \ee
  and the  conformal gravitino \ci{tss,ft3,ftr}
\bea\la{15a}  &&  Z_{{3\ov 2}}  =  \Big[ {      (\det \d_{{1\ov 2}})^5 \ov  ( \det \d_{{3\ov 2}} )^3     }  \Big]^{-1/4}
= ( Z_{{3\ov 2},{1\ov 2}})^3\,    Z_{{1\ov 2}}  =
 ( Z_{{3\ov 2},{1\ov 2}})^2\,    Z_{{3\ov 2},\emptyset} \ , \\ &&
Z_{{3\ov 2},{1\ov 2}} =  \Big[ { (\det \d_{{1\ov 2}})^2 \ov   \det \d_{{3\ov 2}}            }  \Big]^{-1/4} \ ,  \ \ \ \ \ \ \
(\d_{{3\ov 2}})_{ mn}   =  - (\nabla^2_{{3\ov 2}})_{ mn} -  \ha  \g^{kl}  R_{klmn} \ ,   \la{15}\eea
where $Z_{{3\ov 2},\hav}  $ is the partition function of the standard massless     gravitino.
Like for the  Weyl graviton,  the kinetic operator of conformal   gravitino   factorizes   on the
 Ricci-flat background -- here   into  the  product of  the three standard  gravitino operators (with transverse projection)
 and the resulting partition function is equivalent  to the one for 3 massless  gravitino and one massless   spin    1/2    field
 (with total of  8  fermionic  degrees of freedom, cf. \rf{11}).

 As in the bosonic case \rf{12b},  one may  conjecture  that  in general
 the  fermionic   CHS  operator will factorize into the product of   $s$ massless
    (transverse-projected)   operators   $\g^m \nabla_m$ so that  the
  partition function will have again the same  form  as the flat-space one \rf{10f},
 now   with covariant  2nd-order (squared  1st order)    operators $\d_s$  containing only ``minimal'' curvature couplings
 \be\la{inga}
  Z_s =  \Big[ {     (\det \d_{s-1})^{s+1}    \ov    ( \det \d_{s} )^s  }  \Big]^{-{1\ov 2}} \ ,  \ \ \ \ \ \ \ \ \ \ \ \
     s=  {\te \ha, {3 \ov 2}, {5 \ov 2}},  .... \ . \ee
As in the bosonic case,  these $\d_s$  operators  will be  assumed to
have  the form \rf{jo}   acting on
  totally-symmetric $\g$-traceless   real   spinor-tensors $\psi_{m_1...m_\ss}$   corresponding   to the
$ (A,B) \oplus (B,A)$  representation of the Lorentz group with $(A,B)$
contained in  $({\ss \ov 2}  , {\ss\ov 2}) \otimes ( {1\ov 2}, 0)$, i.e.\foot{Here we shall follow
 \ci{cr1,duf}   but not \ci{cr2}:
 the  fermionic higher-spin operator    assumed
in \ci{cr2}   in the $A > B$ case   contained  extra  $1/A$   factor in $X$ in \rf{jo}
   that   may seem somewhat  unnatural   in the context of factorizing higher-derivative CHS  kinetic operators
    (it  may, however, in principle appear upon squaring of 1st-order  fermionic operators, cf. \ci{dow}).
The   operator  in \ci{cr2}
required   strong consistency conditions   that rule out  non-trivial   backgrounds when  applied to real fermions
and thus   do  not allow to compute the coefficient $\b_1$   in the conformal anomaly
for $s > 2$.  Here we  will not worry
about  consistency conditions  of the factor-operators like \rf{jo}  as  such conditions  on the total operator  should be weaker
 in the conformal higher spin  case (and Einstein-space background should be a consistent one).
  Needless to say, the structure of  ``minimal''   fermionic factor-operators in Ricci-flat   background
   for  higher   spins
 $s \geq {5\ov 2}$   deserves further clarification.}
\bea
 (A,B)= {\te ({\ss+ 1 \ov 2}  , {\ss\ov 2})}, \ \ \ \ \  \ \ \ \ \ \ \ \    N_{(A,B)}= (\ss+1 )(\ss+ 2) \ ,\ \ \  \ \ \  s=\ss + \ha \ .
\la{foo}  \eea
Then the corresponding $\b_1$   coefficient is given by the same expression  as in \rf{ww}   up to  an 
overall  minus sign.
Applying \rf{21}  we get    
 \be
&&  \b_1[\Delta_s] = \b_1[\Delta_{(A,B)}]  +  \b_1[\Delta_{(B,A)}] \no \\
&& \ \ \ \ \ \ \ \ \  =
     \te{ 1\ov 2880}  N_s  ( 50 - 28 N_s   + 3 N_s^2  )     \  , \ \ \ \ \ \ \ \   N_s =2N_{(A,B)}=  2 (\ss+1 )(\ss+ 2)   ,  \la{nym}
 \ee
so that  the  $\b_1$ Weyl anomaly  coefficient   for  the fermionic CHS  field  is given by
\be
&&  \b^{(f)}_{1,s} = -    s \b_1[\Delta_s]   +   (s+1) \b_1[\Delta_{s-1}]  \no  \\
&& \ \ \ \ \ =  \te{ 1\ov 2880}  \n_s ( 106 + 90 \n_s + 15 \n_s^2) \  , \ \ \ \ \ \ \ \ \ \ \
  \ \ \n_s = -2 (s+\ha )^2 \ . \la{tak}
\ee
As in the   bosonic case \rf{www},   the anomaly coefficient $ \b_{1,s}$  is again  a  cubic
polynomial in  the number of dynamical degrees of freedom $\n_s$  (see \rf{11}).
For $s= {1\ov 2}$ ($\n_s= -2$)   and $s= {3\ov 2}$  ($\n_s= -8$)
  $  \b_{1s}$ in   \rf{tak}  reproduces the previously  known values   in \rf{46},\rf{47}.

It is   interesting to note that   the expression for $\b_1$   simplifies for   a combination  of CHS fields with spins $(s, s+ \hav, s+1)$
(with integer $s$)  forming an
$\N=1$ supermultiplet:  from \rf{www} and \rf{tak}  one then finds
\be
\b^{(\N=1)}_1 =  \b^{(b)}_{1,s}  +  \b^{(f)}_{1,s+ \hav}    +  \b^{(b)}_{1,s+1}  = \te{ 1 \ov 16} ( s+ 1)^2 ( 4s^2 + 2s -1) \ . \la{coo}
\ee
The  choice of $s=0$ corresponds to $\N=1$  vector multiplet $(\hav,1)$   where   $\b_1= - { 1 \ov 16}$
and $s=1$  to $\N=1$ conformal supergravity  multiplet $(1, {3\ov 2}, 2)$    where $\b_1=  { 5 \ov 4}$,
in agreement with previously known values.





\subsection{Conformally-flat background}

Let us  first recall the  known low-spin cases.
In the case of the conformally-flat   background \rf{5} one finds for $s={1\ov 2},\tri$ partition functions (cf. \rf{14a}--\rf{13})\ci{tss,ft3,ftr}
\be \la{161}  && Z_{\hav}=   \Big[  {1 \ov \det \dn_{\hav } ( 3) }     \Big]^{-1/4}\  , \ \ \ \ \ \ \ \ \
 \dn_{s} ( M^2) = - \na^2_s  + M^2 \ep \ , \\
&&
Z_{\tri}=  (Z_{\tri,\hav})^2  Z_{\tri,\emp} =
 \Big[    {      \det \dn_{\hav} ( -1  )  \ov      \det \dn_{\tri\pe }  ( 3)
     }  \Big]^{-2/4}  \  \ \ \  \Big[       { 1 \ov    \det \dn_{\tri\pe} ( 4)}\Big]^{-1/4}  \la{666}
      \\ && \ \ \ \ \ \  \qquad \qquad \qquad   =
\Big[   {    (\det \dn_{\hav} ( -1  ) )^2 \ \ \ov
   \det \dn_{\tri}  ( 3 ) } \Big]^{-2/4}   \  \Big[ {  \det \dn_{\hav } ( 0)   \ov   \  \det \dn_{\tri} ( 4)      } \Big]^{-1/4} \ .
\la{griv} \ee
Here $\dn_{s}(M^2) \equiv  - \na^2_{s}   + M^2 \ep $  is    defined on
 $\g$-traceless  spinor-tensors,
while $\dn_{s\pe}(M^2)$ is, in addition,
restricted to transverse  spinor-tensors.
The relation   between \rf{666}  and  \rf{griv}  is based on (cf. \rf{ddi},\rf{ddd})
\be
\det  \dn_{\tri} ( M^2 ) = \det  \dn_{\tri\pe } ( M^2 )\ \det  \dn_{\hav } ( M^2 - 4 ) \ . \la{cho}
\ee
One  notes \ci{tss}   that the
  conformal gravitino  partition function \rf{griv}  contains {\it two} factors  of
  the standard ``massless'' Einstein gravitino partition function (with ``cosmological''  mass  parameter
   $\mm^2= - {\Lambda \ov 3} = -\ep$)
\ci{cdgr,fto,ft4}
\be
Z_{\tri,\hav}=  \Big[   {   \det \dn_{\hav} ( -1  )   \ov     \det \d_{\tri\pe} ( 3)      } \Big]^{-1/4}
=   \Big[  {(\det \d_{\hav } ( -1) )^2 \ov     \det \d_{\tri} ( 3)      } \Big]^{-1/4} \la{ei} \ .
\ee
In \rf{666} the $ \det \dn_{\hav} ( -1  ) $   factor  comes from the Jacobian of transformation from
$\psi_m$ to its transverse part  plus pure-gauge gradient part  \rf{a9}
while  $ \dn_{\tri\pe}$ operators   appear   from factorization of the 3rd-derivative   conformal gravitino
kinetic operator in conformally flat  background  \ci{tss,ft3,dn1,dn2}:
\be \la{fag}
\psi_{\tri} \, \hat {\rm D} \, \psi_{\tri} =  \psi_{\tri \pe} \,    \dn_{\tri}  ( 3 )       \,    \hat \na_{\tri}  \, \psi_{\tri\pe}
\ , \ \ \ \ \  \ \ \ \ \ \ \ \ \hat \na_{\tri } \equiv  (\g^k \na_{k})_{\tri}   \ .   \ee
In general, ``squaring''   the ``massive''  1st-order   gravitino
 operator    $\hat \na_{\tri} +   \mm $  
   gives
  \be
    (\De_{\tri})_{mn}
      &=& - (\na^2_{\tri})_{mn}   + \fo R g_{mn}  - \ha \g^{k} \g^l R_{klmn}  + \mm^2g_{mn}\no\\
  &= &- (\na^2_{\tri})_{mn}   + (4 \ep   + \mm^2) g_{mn}  \ , \la{res}
  \ee
  where in the second line we  assumed the  conformally-flat background \rf{a1}.\foot{Note that  in this paper we use  different notation compared to  \ci{tss,ft3,ftr}:    there  this operator was  denoted as $  \De_{\tri\pe} (\mm^2)$   while here it is
   called $\dn_{\tri\pe}  (4    +  \ep^{-1} {\mm^2})$.
   Similarly, $\De_{\hav} (\mm^2) = - \na^2_{\hav}   + \fo R + \mm^2 = - \na^2_{\hav}   + 3 \ep  + \mm^2$
   here is  called  $\dn_{\hav}  (3    + \ep^{-1} {\mm^2})$.
}
Thus  the   second power   of   $\dn_{\tri\pe}  ( 3 ) $  (with  $\mm^2 = -\ep$)
 in \rf{666}  has to do with its   appearance
already in the original fermionic action,   while the extra   factor
  $\dn_{\tri\pe}  ( 4 ) $    comes from squaring
of the $ \hat \na_{\tri} $  operator  in \rf{fag}, i.e.
  \rf{res} with $\mm=0$.
The fact that this  ``extra''   operator is just the  square of the  standard  ``{\it mass-zero}''
 transverse $\g$-traceless
gravitino operator  $\hat \nabla_{\tri}$  explains  also its  special conformal-invariance
property  (implying ``null-cone'' propagation) \ci{dn1,dn2}.
Thus  this  operator is  a  direct  counterpart of
$\dn_{2\perp} (4 )$  in the  conformal graviton $s=2$  case   \rf{wey},\rf{133}    but lacking
extra effective   gauge  invariance (reflected in the trivial numerator in the second factor of \rf{666})
 the corresponding field  was not called  ``partially massless''  in \ci{dw1,dw2}
 and it describes a  massive  state (with $2s+1=4$  dynamical  d.o.f.)  in (A)dS$_4$.\foot{Let us recall that the definition of ``mass''
 is ambiguous in (A)dS   and  truly   massless field (with 2 degrees of freedom)    corresponds  to maximal amount of gauge invariance
 (and thus smallest number of propagating modes).
 PM   fields  have less gauge invariance (and thus more degrees of freedom),  with generic massive fields having no
 residual gauge invariance.
}

The above discussion  suggests the following natural generalization of the flat-space
fermionic CHS  partition function \rf{11f} to the conformally-flat Einstein-space case
which  is   a  direct counterpart of the bosonic   expression  \rf{fi}.
We shall assume that  for all $s\geq { 3 \ov 2} $ the $2s=2\ss+1$-derivative  fermionic CHS  field kinetic  operator
factorizes, like in the bosonic case \rf{pass},
into the product  of ``squares'' of  all $\ss$  ``partially massless'' (PM)    1st-order fermionic  spin $s$ operators
$\hat \nabla_s  + \mm_{s,k}$ with special mass parameters
 $\mm_{s,k}$  ($k=\hav,...., s-1$)
 and also  one extra ``mass-zero''
operator $\hat \nabla_s $ (which, in fact, represents a massive state in (A)dS)
 \be
  \psi_s {\rm D}_s \psi_s =
  \psi_{s\pe}  \Big[ \prod_{k=1/2}^{s-1} ( \hat \nabla_s  + \mm_{s,k} )( - \hat \nabla_s  + \mm_{s,k} )         \Big]
\   \hat \nabla_s \  \psi_{s\pe} \  , \ \ \ \ \ \ \ \ \ \ \ \ \ \  \hat \na_{s} \equiv ( \g^k \na_{k})_s \ .  \la{fass} \ee
As in the bosonic case \rf{pass},
this factorization is   suggested, in particular,   by  the existence of  an ``ordinary-derivative'' formulation
 of the fermionic CHS fields   \ci{metcf}.

The values of the fermionic PM   mass
parameters (first conjectured in $D=4$ in \ci{dw2} and   confirmed and extended to any $D$ in
\ci{mef}) are
\be \la{f12}
\mm^2_{s,k} =  -  ( k + \ha  + {\te { D-4 \ov 2}} )^2\ep \ , \ \ \ \ \ \ \
\mm^2_{s,k}\Big|_{D=4}  =  -  ( k + \ha )^2\ep \ , \ \ \ \ \ \ \  k=\ha, ..., s-1\ ,
\ee
where $k=s-1$   corresponds to the massless field in (A)dS$_D$.
The  PM  fields  admit  a local  gauge-covariant description  upon introducing extra
lower-spin  fields \ci{mef}; eliminating the latter   gives  residual  gauge transformations
with higher $\nabla^{s-k}$ derivatives acting on lower-rank spinor-tensor parameters $\xi_{k}$.

Starting with  generic operator $\hat \nabla_s  + \mm$  describing
 massive  on-shell   spinor-tensors ($s=\ss+\ha$)
\be
( \gamma^k  \nabla_k +   \mm ) \psi_{m_1 ... m_\ss} =0 \ , \ \ \ \ \ \  \ \ \
\g^{m_1} \psi_{m_1 ... m_\ss} =0 \ , \ \ \ \ \ \ \  \nabla^ {m_1} \psi_{m_1 ... m_\ss}  =0 \ ,
\ee
 and ``squaring''  it  gives in conformally-flat   case  the following  operator
  (generalizing  the $\ss=1$ one in \rf{res})\foot{A
 derivation  of the $( \ss + 3) \ep $   contribution to the mass  term
 can be given, e.g.,
by  considering $X=- \ha \g^m \g^n [\nabla_m,\nabla_n] =   - \ha \g^{m} \g^n  R^{ab} _{\ \ mn }  \hat \Sigma_{ab} $
 where $\hat \Sigma_{ab}$ corresponds to the representation describing  spinor-tensor $\psi_{m_1 .. m_\ss}$
 (generalizing $\ss=1$ case in \rf{res}).
 }
 \be \la{sq}
 \dn_s (M^2)   = ( - \hat \nabla_s +   \mm ) ( \hat  \nabla_s +   \mm )
 = - \na^2_s   + \MM^2  \ , \ \ \ \ \ \ \
 \MM^2 = (\ss+3) \ep + \mm^2 \equiv M^2 \ep  \ .
 \ee
 Thus the  family  of PM fermionic operators in $D=4$  is represented by
 the following set  of  2nd-order operators   (cf. \rf{pm})
 \be \la{pmf}
&& \dn_s (M^2_{s,k})   = - \na^2_s   + M^2_{s,k}  \ep \ , \ \ \ \ \ \ \ \ \ \ \
M^2_{s,k} = \ss + 3     -  ( \kk + 1)^2 \ ,\\ &&  \ \ \ \ \ \
\ss= s- \ha=0, 1, ...\ , \ \ \ \ \ \  \ \ \qquad  \kk= k - \ha=
\kk=0, ..., \ss-1 \ .\no  \la{ppf} \ee
 $k= s-1$    corresponds to  the   standard massless  case  (here $D=4$, cf. \rf{ppm})
\be
 m^2_{s0}\equiv M^2_{s,s-1}= \ss + 3   -  \ss^2 =   2 s  - s^2   + {\te{9 \ov 4}}  \ , \la{pl} \ee
 which is the only  choice for $s=\tri$ ($\ss=1$)  case.
 The first  non-trivial  PM  field appears   for $s=\fiv$
 where we   get  for $\kk=1$ and $\kk=0$:
\ $
 M^2_{\fiv,\tri}=   1      , \ \    M^2_{\fiv,\hav}=  4  $.

 One  extra  ``genuinely-massive'' operator  that   we should add
 corresponds   to $\mm^2=0$ in \rf{sq}:  it  can be viewed    as  a natural member of an  ``extended'' PM
 family \rf{pmf}
 were we allow   also the $\kk=-1$ ($k= -\ha$)  value:
\be \la{pmff}
  \dn_s (M^2_{s,\emptyset})   = - \na^2_s   + M^2_{s,\emptyset}  \ep \ , \ \ \ \ \ \
\ \ \  \ \ \  M^2_{s,\emptyset} \equiv M^2_{s, -\hav} =  \ss + 3   \ . \ee
 As in \rf{11f}  here the index  $\emptyset$ indicates that there is no associated gauge invariance, i.e.
 this field  describes $2s+1$  degrees of freedom  \rf{11ff}.
 The  set of  such  (A)dS-massive  but conformally-invariant fields
  includes  the standard  fermion  \rf{161}  ($\ss=0$)
 and the $\dn_{\tri\pe} (4)$  gravitino in \rf{666}  ($\ss=1$).

 As a result,  the   fermionic CHS partition function in conformally-flat background
 should   have the  following  representation (that directly reduces to \rf{11f} in the flat-space $\ep=0$ limit)
   \be  && Z_s =\prod_{k=\hav  }^{s-1} (Z_{s,k})^2 \   Z_{s,\emptyset} \ , \la{2ff}\\
   &&
 Z_{s,k}=   \Big[  { \det  \dn_{k  \pe } (M^2_{k,s})   \ov  \det \dn_{s\, \pe} (M^2_{s,k}) }\Big]^{- 1/4}  \ ,   \ \ \ \ \ \ \ \ \ \ \ \
  Z_{s,\emptyset}=    \Big[ { 1 \ov \det \dn_{s\, \pe}  (M^2_{s,\emptyset})  }\Big]^{- 1/4} \ .
 \la{22f}
 \ee
 Here we  used that as in the corresponding bosonic expression
  in \rf{fi}  the part of the Jacobian of transformation
 from $\psi_s$  to $\psi_{s\pe}$  and  other low-rank  reducible components  that remains  after the division over the volume of    gauge    group is given by
 \be \la{jac}
  \prod_{k=\hav  }^{s-1}  [\det  \dn_{k  \pe } (M^2_{k,s})]^{-1/2}
 \ , \ \ \ \ \ \ \ \ \       M^2_{k,s} = \kk + 3     -  ( \ss + 1)^2 \ , \ee
    where  $M^2_{k,s}$   is again   the ``transpose''   of the PM mass  matrix    in \rf{ppf}.
    Let us note also that   for half-integer   $k$ one   has the following counterpart of \rf{ddi}
    \be   \la{rat}
 \det \dn_{k\perp}(M^2)   = { \det \dn_{k}(M^2 ) \ov   \det \dn_{k-1}  ( M^2- \delta_k) } \ , \ \ \  \ \ \ \ \ \
    \delta_k\Big|_{D=4} = 2 (\kk +1)  \ ,
\ee
which generalizes  \rf{cho}.

 As already   mentioned,  the $k=s-1$  factor in \rf{2ff}
  is the square of the standard massless  spin $s$ partition function in (A)dS$_4$
    (cf. \rf{mas},\rf{mast} and \rf{ei})
    \be
    Z_{s,s-1} =  \Big[  { \det  \dn_{(s-1)  \pe } (M^2_{s-1,s})   \ov  \det \dn_{s\, \pe} (M^2_{s,s-1}) }\Big]^{- 1/4}
    =  \Big[  { \det  \dn_{(s-1)  \pe } (1-\ss-\ss^2)   \ov  \det \dn_{s\, \pe} (  3 + \ss - \ss^2) }\Big]^{- 1/4}\ .
   \ee
 The special cases of \rf{2ff}  for $s=\hav$ and $s= \tri$  of course agree  with \rf{161} and \rf{666}
 while, e.g.,   for $s=\fiv$ we get
 \be
 Z_{\fiv}&=&  (Z_{\fiv,\tri})^2  (Z_{\fiv,\hav})^2  Z_{\fiv,\emp}\no   \\  &=&
 \Big[    {      \det \dn_{\hav} ( -5  )  \ov      \det \dn_{\fiv\pe }  ( 1)
     }  \Big]^{-2/4}  \   \Big[    {      \det \dn_{\hav} ( -6  )  \ov      \det \dn_{\fiv\pe }  ( 4)  
     }  \Big]^{-2/4}  \ \  \Big[       { 1 \ov    \det \dn_{\fiv\pe} ( 5)}\Big]^{-1/4}   \la{555}\ .
\ee
To  compute the   Weyl-anomaly coefficient $\aa_s$  corresponding to \rf{2ff}  we start again with the 
general relations \rf{h1},\rf{f1}  applied now  to the  case of the representation \rf{foo}.
Then the counterparts  of \rf{aa2},\rf{aa22}   in the half-inter spin $s$ case  are (see \rf{rat})
\be  &&  \aa[ \dn_s (M^2)] =  - \te{ 1 \ov 144}    N_s   \Big(  N_s  -   3 M^4
  + 12 M^2   - {\te { 121 \ov 10 }}    \Big) \  , \ \ \ \ \ \    \ \ \  N_s = (\ss+1) (\ss+2)\ ,  \la{af2}  \\
    &&  \aa[ \dn_{s\pe} (M^2)] =  \aa[ \dn_{s} (M^2)] -  \aa[ \dn_{s-1} (M^2 - 2\ss-2  )]
\no \\ && \ \ \  \ =
- \te { 1 \ov 720}   (\ss+1)    \Big[  -101 + 20 \ss ( 3 \ss^2 + 13 \ss  + 11)     - 60  (\ss^2 + \ss  -2) M^2     - 30  M^4     \Big] \ . \la{af22}
\ee
Here  we already accounted for an extra $1\ov 2$ factor  and fermionic minus sign, i.e.,
for example,  $  \aa[ \dn_{\hav} (3)]  = { 11 \ov 720}$ gives the contribution of a single  $s=\hav$   fermion in   \rf{161}.
Applying this  to operators  in \rf{22f}  with mass parameters given in \rf{pmf},\rf{pmff},\rf{jac}
and performing the sum over $\kk=0, ...,\ss-1$  as required by \rf{2ff}   we end up with the following expression for the
fermionic  CHS $\aa_s$-coefficient  which is a  counterpart of the bosonic expression in \rf{nen}:
\be
\aa^{(f)}_s&=&
2 \sum_{\kk=0}^{\ss-1}
  \Big(   \aa[ \dn_{s\pe} (  2 + \ss   - 2 \kk -\kk^2    )]     -        \aa[ \dn_{k\pe} ( 2 + \kk  - 2\ss - \ss^2    )] \Big)
+ \aa[ \dn_{s\pe} ( 3 +  \ss   )]
 \no\\
 &=& \te { 1 \ov 2880} \nu_s  ( 12 + 45 \nu_s   + 14 \nu_s^2 ) \ , \ \ \ \ \ \ \ \ \ \ \ \ \ \ \ \ \ \ 
   \nu_s = -2 (s+\hav)^2      \ . \la{ahf} \ee
  In particular,
  $\aa^{(f)}_{\hav} = { 11 \ov 720}, \ \aa^{(f)}_{\tri} =- { 137\ov 90}$ in agreement  with \rf{46},\rf{47},
   also
  $\aa^{(f)}_{\fiv} =- { 1869 \ov 80},$ etc.

Combining the results   for $\b^{(f)}_{1,s}$ in \rf{tak}   and $\aa^{(f)}_s$ in  \rf{ahf}   we  conclude  that
\be
\cc^{(f)}_s = \ha \b^{(f)}_{2,s}= \b^{(f)}_{1,s} + \aa^{(f)}_s
 =  \te{ 1 \ov 2880} \n_s (118   +135  \n_s  +   29  \n^2_s)  \ . \la{nin}
\ee
Like the bosonic expression in \rf{nen}, the    cubic polynomial  \rf{ahf} in $\nu_s$  turns   out to be
special:   when summed over all  spins $s= \hav, ..., \infty$   and analytically ($\z$-function) regularized it gives
 zero. This will not be true, however, for  the sum of $\cc^{(f)}_s$ in \rf{nin}.
  We shall discuss  this   in more detail  in the next section.

  \renewcommand{\theequation}{5.\arabic{equation}}
 \setcounter{equation}{0}
\section{Concluding remarks}

Our final results for the Weyl anomaly coefficients \rf{1} of $D=4$ conformal higher spin  fields   were
already summarized in \rf{19}--\rf{20f}.
As was mentioned  in the Introduction,  ref.\ci{gio}    made a   remarkable observation  that the sum 
over all spins of the bosonic $\aa_s$ coefficient in \rf{19}  gives zero
 if computed  using $\z$-function prescription,
suggesting the   existence of  an anomaly-free theory.\foot{Here we  consider   just on  the conformal 
spin 2  gauge symmetry preservation:   the reparametrization
  invariance  should be manifest, while the Weyl  invariance anomaly should cancel out.
 Anomalies of  higher spin analogs of these $s=2$ symmetries  (in particular, higher spin trace anomalies,  cf.\ci{bek,ruh}) should also  be absent  for the full consistency
 of the theory.} 
  The same    happens to be true also for the
corresponding sum in the fermionic  case (see \rf{fif}).

Let us   now discuss this   vanishing  of the regularized sum of $\aa_s$ anomaly  coefficients  in more detail.
As is well known, to  define a power-divergent sum like $P_n= \sum_{s=0}^\infty  p_n(s)$  where $p_n$ is an order $n$ polynomial in $s$
one should not, in general, use a sharp cutoff (like   $ 0 \leq  s \leq M$, $M\to \infty$)  but  should
consider a smooth analytic regularization  with a cut-off function $f$, i.e. define
 (see,  e.g., \ci{brink} or \ci{bil} for a recent discussion)
 \be P_n\Big|_{\rm reg} = P_n(\ep\to 0)\Big|_{\rm fin}\ , \ \ \ \ \ \ \
P_n(\ep) \equiv  \sum_{s=1}^\infty     f(\ep s) \ p_n(s)\   , \qquad  \ \ \ \ f(0) =1  \ , \ \ \ \ \    f(\infty) =0  \ , \la{eer}
\ee
one should  compute the regularized sum,  take the limit  $\ep\to 0$  and drop  all singular $ 1\ov \ep^m$ terms.
  For example, one may use  an exponential cutoff
 $f(\ep s) = e^{ - \ep s} $. Then $(P_n)_{\rm reg}$ will be the same
 as found  by  computing each term $\sum_{s=1}^\infty  s^k$   using the $\z$-function regularization. 

 One   may wonder what is the  physical meaning of this   regularization prescription in the present   ``sum over spin'' context.
 A possible  answer  is  that it  is required to preserve  some hidden symmetries  of the    higher-spin system  (cf.  \ci{gk}).
 In fact, one can draw an analogy   with string theory which describes  an infinite set of fields  of   growing  spins   and masses
 which are effectively summed over in the  world-sheet   description.  Indeed, a standard example is  that  the
 use of  an  analytic  or $\z$-function  regularization   of oscillator sums in computing, e.g.,
  2d central charge and  vacuum energy
 \ci{brink}  gives,  for example,  the right  (zero) value for the  mass of the first excited level   state
 in bosonic open string (photon) and is thus required   for a consistent realisation  of target space symmetries  of  string  theory.

 Computing  the  sum of  $ \aa^{(b)}_s $ in \rf{19}  using the  exponential cutoff we get
 \be &&
\sum_{s=1}^\infty\,   e^{- \ep s} \,   \aa^{(b)}_s  = \te
 \frac{e^{2 \ep} (566 e^{\ep}+1326 e^{2 \ep}+566 e^{3 \ep}+31 e^{4 \ep}+31)}{180 (e^{\ep}-1)^7} \no \\
&&\qquad \qquad \ \
 = \te { \frac{14}{ {\ep}^7}+\frac{7}{{\ep}^6}+\frac{3}{2 \text{\ep}^5}+\frac{1}{6 \text{\ep}^4} +\frac{1}{120  \text{\ep}^3}   +  { \ep \ov 7560}  + O(\ep^2) }\ .
 \la{gf1} \ee
 Thus   the finite  part of this sum vanishes as claimed in \rf{fif}. The same result is, of course,
  found \ci{gio} using $\z$-function regularization.\foot{Computing    $P_n(\ep) = \sum_{s=0}^\infty   e^{-\ep s}\,  s^n $,
    and dropping all singular terms  in  $\ep \to 0$  one gets
  the finite part  $P_n\big|_{reg}=\z(-n)$. Explicitly,  for odd $n$:
 $P_1= {1\ov \ep^2} - {1\ov 12}   + O(\ep^2)$,
 $P_3 = {6\ov \ep^4}  +  {1\ov 120}   + O(\ep^2)$,
 $P_5 = {120\ov \ep^6} - {1\ov 252}   + O(\ep^2)$, etc.,
 while  for even $n$:
 $ P_0= {1\ov \ep} - {1\ov 2}   + O(\ep)$,
 $P_2= {2\ov \ep^3} - {\ep\ov 120}   + O(\ep^2)$,
 $P_4 = {24\ov \ep^5}  + {\ep\ov 252}   + O(\ep^2)$, etc. }
Similarly, in the case of the fermionic $\aa_s$ coefficient  in \rf{19f} we get
\be
\sum_{s=\hav}^\infty   e^{- \ep s} \,  \aa^{(f)}_s
 &=&  \sum_{\ss=0}^\infty   e^{- \ep (\ss + {1\ov 2} ) }\,  \aa^{(f)}_s
\no\\
&=&\te \frac{e^{\frac{3 \text{\ep}}{2}} \left(-1173 e^{\text{\ep}}-8918 e^{2 \text{\ep}}-8918 e^{3 \text{\ep}}-
1173 e^{4 \text{\ep}}+11 e^{5 \text{\ep}}+11\right)}{720 (e^{\text{\ep}}-1)^7}\no\\
&=&\te {
-\frac{28}{\ep^7}-\frac{14}{\ep^6}-\frac{2}{\ep^5}+\frac{1}{6 \ep^4}+\frac{47}{480 \ep^3}+\frac{1}{64 \ep^2}
+\frac{7}{5760 \ep}
+\frac{3607 \ep}{7741440}
+O(\ep^2)\  }
\la{fif1}
\ee
so that   again there is no left-over  finite part.
Note that here the  original sum   goes over half-integer spins,  so that to apply the  equivalent
 $\z$-function regularization prescription one needs  to use
 $\z(z, q) = \sum_{n=0}^\infty ( n+q) ^{-z}$ with $q=\hav$, i.e.
$\z(z, \hav) = ( 2 ^z -1) \z(z)$.\foot{This is  again similar to the
prescription one uses  in string theory   when computing, e.g., the   vacuum energy
 in  NS sector   where  2d fermions are anti-periodic.
Explicitly, the cancellation of the finite part in \rf{fif1}  is due to the following relations:
$\z(0,\hav)=\z(-2,\hav)=\z(-4,\hav)=0$ and ${ 1 \ov 64} \z(-1,\hav) + { 1 \ov 36}  \z(-3,\hav)- { 7 \ov 60}  \z(-5,\hav)=0$.}

Let us stress   that this    vanishing   of the regularized   sums of
the  $\aa_s$-coefficients  is    non-trivial.
 Together     with the known  lower spin  results  for $\aa_s$
in  \rf{44}--\rf{47}    this property  uniquely fixes the expression for $\aa_s$  in both   the  bosonic and  the  fermionic cases.
First, one may argue on general grounds (from the structure of $b_4$ in \rf{algo}   and the
form of the partition functions \rf{fi},\rf{2ff}) that the  conformal anomaly coefficients $(\aa_s,\cc_s)$
 should be   given by cubic homogeneous polynomials
of $\n_s$, i.e. of the physical  number of d.o.f. of a spin $s$ field. 
In the bosonic case demanding agreement with the known  $s=1,2$  values in \rf{44},\rf{45}  then   leads to the following predictions
\be
 &&  \aa^{(b)}_s=  { \te{ 1 \ov 720}} \nu_s \Big[ \nu_s ( 3 + 14 \nu_s)  \ \ \ \ +\ \     q^{(b)} ( \n_s -2) (\n_s-6) \Big]        \ , \la{ah2} \\
  &&  \cc^{(b)}_s=  {\te{ 1 \ov 1080}} \nu_s\Big[ \nu_s( -59 + 43  \nu_s)    + \   r^{(b)}  ( \n_s -2) (\n_s-6) \Big]          \ ,
   \ \ \ \ \ \   \n_s = s(s+1) \ . \la{ch2}\ee
Similarly,   in the fermionic case  the expressions that match
 the known $s= \hav, \tri$  values in  \rf{46},\rf{47}    are
 \be
  &&  \aa^{(f)}_s=  { \te{1 \ov 23040 }}\  \nu_s \Big[ \nu_s ( 300+ 106  \nu_s)   +    q^{(f)}   ( \n_s +2) (\n_s+8) \Big]
       \ , \la{fh2} \\
  &&  \cc^{(f)}_s=  { \te{1 \ov 23040}}\  \nu_s\Big[ \nu_s( 490  + 173  \nu_s)    +     r^{(f)}   ( \n_s +2) (\n_s+8) \Big]           \ ,
   \ \ \ \  \nu_s= -2 ( s + \ha)^2\ .     \la{ff2}
  \ee
  Here  $ q^{(b)},r^{(b)}$   and $ q^{(f)},r^{(f)}$  are so far arbitrary  coefficients.
Now imposing     the  additional condition   of  the vanishing  of finite parts  the   corresponding
  sums of $\aa_s$ over all spins (integer in the bosonic case and half-integer in  the fermionic case)  
  fixes  (after computing the sums as   in  \rf{gf1}, \rf{fif1})  the coefficients
    $ q^{(b)}$ and $ q^{(f)}$ uniquely
\be  \la{un}     q^{(b)}=0 \ , \ \ \ \ \ \ \ \ \ \ \ \ \     q^{(f)}=6  \ .  \ee
Then \rf{ah2} and \rf{fh2}   become  precisely  to  the expressions  for $ \aa^{(b)}_s$ \rf{nen}  and 
  and $ \aa^{(f)}_s$  \rf{ahf} that were independently found  above    from the
  detailed structure of the CHS partition functions  \rf{fi} and \rf{2ff}
in (A)dS$_4$  background.

Applying the same requirement  of  zero  finite part   to the  regularized
sums   of  the $\cc_s$-coefficients in \rf{ch2},\rf{ff2}, i.e.
 $\sum_{s=1}^\infty  e^{-\ep s} \, \cc^{(b)}_s$   and
$\sum_{s=\hav}^\infty e^{-\ep s} \,   \cc^{(f)}_s$,    gives
\be  \la{un2}    r^{(b)}_{0}  = -1    \ , \ \ \ \ \ \ \ \ \ \  r^{(f)}_{0}  =  \te{ 3597 \ov 367}     \ . \ee
The results  for $\cc^{(b)}_s$ in  \rf{nyn}  and $\cc^{(f)}_s$  in \rf{nin}  that we have found   above
correspond,  however, to  different values
\foot{It  may be  interesting to note that the
 difference between  our  value   of $\cc^{(b)}_s $ with  $r^{(b)} =\ha $
and the  ``zero-sum'' value  with   $ r^{(b)} = -1$ is  an integer (a  binomial coefficient)
\be \la{cec}\no
 \te { \cc^{(b)}_s \Big|_{r^{(b)} ={1 \ov 2}}  - \cc^{(b)}_s\Big| _{r^{(b)} =-1} = {1 \ov  720  }{ \nu_s ( \n_s -2) (\n_s-6)}  =
 {\te { 1 \ov 6!}} s ( s^2-1) (s^2 -4) (s+3) =
 {\te {\pmatrix{ s + 3\cr 6}} }. }
 \ee
 }
  \be r^{(b)} =\te   { 1 \ov 2}     \ , \ \ \ \ \ \ \ \ \ \  r^{(f)}  = 59 \  .
  \ee
  This  non-vanishing  of sums of $\cc_s$  we have found suggests  that the expressions in \rf{nyn}  and  \rf{nin}
may   deserve further  checks.


Let us note also that the   vanishing   \rf{fif} of the regularized sums of the $\aa_s$ coefficients \rf{gf1},\rf{fif1}
means also the UV finiteness  of the  products of the (A)dS partition functions:
$\prod_{s=1}^\infty Z^{(b)}_s$   corresponding  to   \rf{fi}  and
$\prod_{s=\hav}^\infty Z^{(f)}_s$ corresponding  to    \rf{2ff}.

\

Here we  discussed  only the $b_4$ heat kernel coefficient   of the   logarithmically divergent part
of   CHS free   energies $\ln Z_s$  but
the  corresponding   partition functions  on  (A)dS$_{D}$ or $S^D$   may be computed explicitly as, e.g., in \ci{alle,fto,camp,gio}.
This   should  allow one to prove directly
the relation between,   e.g.,  the   bosonic   conformal higher spin $s$
partition function \rf{fi}  on $S^4$   and  the ratio of  partition functions of  massless  spin $s$ field in AdS$_5$
with alternate boundary conditions
as implied by the  AdS/CFT in the context of   ``double-trace'' deformation   construction    \ci{gio}.

Let us briefly review some underlying ideas.  Coupling, e.g.,  the  $D=4$
conformal  $\N=4$  SYM theory  to a  background
conformal supergravity multiplet and integrating out the SYM fields
one finds an  induced  action  for the conformal supergravity fields  \ci{ft2,lt,buc}:
$ S_{\rm eff} \sim   \int  C_{mnkl}   \ln (L^{-2} \nabla^2 )  C_{mnkl} + ...  \sim   \int  (  C^2_{mnkl} + ...)+ ${ non-local} terms.
The quadratic and cubic terms in this action expanded in powers of the
 fields summarize information about the protected  2- and 3-point  SYM correlators like
$\langle  T_{mn} T_{kl}\rangle $ and $\langle  T_{mn} T_{kl} T_{sr}\rangle $.
The ``protected''  part  of this induced   action 
(which is the same at strong and weak coupling) appears also upon  solving the  Dirichlet problem
in the 5-d  $\N=8$ gauged supergravity on the AdS$_5$ background.
This relation  can be generalized \ci{tse}  by  starting   with the  free  $\N=4$ gauge  theory   and coupling
 it to a
higher spin  generalization of the conformal supergravity multiplet.
 Let us consider, e.g., the  bosonic  conserved traceless bilinear currents \ci{ber} 
  $J_{m_1...m_s}\sim  X_r \del_{n_1} ... \del_{n_s} P^{n_1...n_s}_{m_1 ...m_s}   X_r $  (cf.\rf{7}; $X_r$ stand for the CFT fields)
  of    dimension $\Delta= 2 + s$.\foot{Note that in general  the mass dimensions of  different fields  involved are
  (we assume that $D$ is even):
  boundary scalars with action $\int d^D x\  X_r \del^2  X_r$:   \ $\Delta = \ha( D-2)$;
  conformal current $J_s\sim  X_r \del^s  X_r$:    \ $\Delta= s + D-2$;
  the corresponding ``source''  field  ($\int d^D x\ J_s \p_s$)   --
  conformal field  with action $\int d^D x\ \p_s  P_s \del^{2s+D-4} \p_s$: \ $\Delta= 2-s$.}
 Coupling them  to  a    background higher spin conformal  field $\p_s$,
integrating out the free SYM fields 
 and expanding the resulting induced effective action  for $\p_s$
 to quadratic order  one  then gets  the logarithmically divergent term proportional to the
 CHS   Lagrangian $\int d^4 x\ \p_s P_s  \del^{2s} \p_s$. It  can  be matched with the term
  originating from  the classical  free action of the corresponding ``dual''   higher spin  massless field $\vp_s$
   in AdS$_5$
evaluated on the solution of a  Dirichlet problem with $\p_s$   as the  boundary data.
As in the $s=2$  case of conformal  (super)gravity  multiplet,  this
agreement between the free bulk massless higher spin  action
and  the  induced boundary conformal higher spin action   is  essentially kinematical, i.e.  is
 guaranteed  by  symmetries (see also \ci{vvv})
 and  applies,  of course,  not only  in  $D=4$    but also  in  other dimensions (see 
 \ci{mett,bek,bek2,gio}).
  
In addition to this ``tree-level''  relation  between free action of massless   higher spins on AdS$_{D+1}$ and free action of
conformal higher spins  on   S$^D$ there is also a ``one-loop'' relation \ci{gio} motivated by the AdS/CFT correspondence in  the
presence of the ``double-trace'' deformation \ci{rast,guk,har,dorn}.
Namely,  the 1-loop  determinant of the CHS  kinetic operator on S$^D$ should be  equal to the  ratio of the massless  higher spin 1-loop determinants in euclidean AdS$_{D+1}$  with alternate boundary conditions.
This ``one-loop''  relation is  more   subtle  than the  ``tree-level'' one  mentioned above
but it should  be   possible to prove it   directly
by  comparing the corresponding heat kernel representations (cf. \ci{har,dorn}), now that
 the expression for the CHS  partition function in terms of the standard 2nd-derivative operator determinants on S$^D$ is known  \rf{fi}.

To motivate this relation one   starts with a large $N$  CFT
 free energy $F=-\ln Z$   on S$^D$  and considers   its change upon
 RG flow from UV  to IR    induced by ``double-trace''  $ \g ( J_s )^2$ deformation.
This  change corresponds  to  alternate $\Delta_\pm $   boundary  conditions
 for a massless  higher spin in AdS$_{D+1}$.
Considering,  e.g.,  as a   boundary CFT   $N$  free scalars on S$^D$
  and introducing an auxiliary field $\p_s$ one  may replace  $\g ( J_s )^2 \to    J_s \p_s -  {\te { 1 \ov 4 \g}} ( \p_s )^2$.
   The resulting ratio of  large $N$ partition functions with and without  the  ``double-trace''
   deformation   is given by the path integral over $\p_s$ with the action
    being the induced effective action for $\p_s$   found by integrating out the original  CFT scalars,
   $ N  \int \p_s  P_s \del^{2s+ D-4}  \p_s$
  plus the $( \p_s )^2$  term.
   Then for large $N$  the  latter can be ignored   and  the leading-order
   result should is given just by the 1-loop partition function of the conformal higher spin  field on S$^D$.

Given  a  generic  massive spin $s$ field  equations in AdS$_{D+1}$ 
    with the transverse traceless kinetic operator
  \rf{33}  with $M^2 = m^2_{s0} + m^2$  (with $m^2_{s0}$  given by
  \rf{ppm} with $D\to D+1$) one finds the solutions behaving near the  AdS boundary ($z\to 0$)
  as $\vp_s \sim   z^\delta, \ \delta= {\Delta - s}$,  where  $[\Delta - (2-s) ] [\Delta - (s + D -2)]
   = m^2$, i.e. $\Delta_\pm = \ha D \pm [ (\ha D + s -2)^2 + m^2 ]^{1/2}$ 
   \ci{met}.
   These two  values of  dimensions   correspond   to the dimensions of $J_s$ in the
   two boundary CFT's  
    which are the end-points of the RG flow induced by the ``double-trace'' deformation \ci{gio}
    ($\Delta_+$   corresponds to  the original free IR CFT and $\Delta_-$ to the    UV   CFT). 
  In the $m=0$   case the two values  $\Delta_\pm$  are  equal to
  the dimensions of the conserved  current $J_s$ ($\Delta_+=s + D-2$)  and $\p_s$  ($\Delta_-=2-s $).

From the AdS$_{D+1}$  theory side   the corresponding  order $N^0$ term in the partition  function   should be given
by the 1-loop  partition function of the  AdS  massless spin $s$  field  with the  appropriate boundary conditions.
One is then  led     to the following  relation  between
 the ratio  of the  1-loop  massless   higher spin  AdS$_{D+1}$ partition functions  evaluated with alternate $\Delta_\pm$   boundary conditions
and the conformal higher spin partition function in S$^D$ 
\be
  &&  {Z^{(-)}_{s0} \ov Z^{(+)}_{s0} }\Big|_{AdS_{D+1}}  =  Z_s\Big|_{S^D}  \ , \la{zz}
 \\
 &&  Z_{s0}\Big|_{AdS_{D+1}} =
 \Big[{\det \big[-\nabla^2  + (s-1) ( s + D-2) \big]_{s-1\, \pe} \ov
 \det \big[-\nabla^2  -  s + (s-2) ( s + D-2) \big]_{s\pe}}\Big]^{1/2} \ , \la{zza}\\
 &&  Z_{s}\Big|_{S^D} =\prod_{k=0}^{s-1} \Big[{\det \big[-\nabla^2   + k - (s-1) ( s + D-2) \big]_{k\, \pe} \ov
 \det \big[-\nabla^2  + s - (k-1) ( k + D-2) \big]_{s\pe}}\Big]^{1/2}
 \no \\
&& \qquad\  \ \ \ \  \  \times 
 \prod^{-1}_{k'=-{1\ov 2}({ D-4 } )} \Big[{1 \ov
 \det \big[-\nabla^2  + s - (k'-1) ( k'+  D -2) \big]_{s\pe}}\Big]^{1/2}
   \ . \la{azz}
\ee
$Z_{s}$  is the CHS partition function on S$^D$  given  in  \rf{fi},\rf{pm},\rf{maa}
(we set the  radii to 1, i.e.  $\ep_{\rm AdS} =-1, \ \ep_{\rm S} = 1$).
In $Z_s$ we   included 
the   extra ``massive''  factor \rf{exx},\rf{ex11}   present for $D > 4$. 
 $ Z_{s0} $  is the massless  spin $s$ partition function in  AdS$_{D+1}$
(given  by    \rf{mas}   with $D\to D+1$).\foot{As explained in \ci{gio}, the
choice of  the boundary conditions for ``ghost'' determinant in $ Z_{s0}\big|_{AdS_{D+1}} $
($\xi_{s-1} \sim z^{\delta_\pm}$, \ $\delta_+ = D, \ \delta_- = 2 - 2 s $) is  correlated with the
$\Delta_\pm$ choice for the physical operator.}


To summarize, 
an  argument leading to   \rf{zz}  includes the following steps (see \ci{guk,dorn,gio}): 
 (i)   one starts with the partition function $Z(\g)$  of   $D$-dimensional 
 UV  CFT  deformed by  a relevant operator   $\g J_s^2$   where   $J_s$ has dimension 
 $\Delta_- = 2-s$;  (ii) one ``splits''  $J_s^2$  as  $J_s \p_s$ plus $\g^{-1} \p_s^2$ 
  introducing   an auxiliary field $\p_s$   of dimension $\Delta_+ =  D - \Delta_-$; 
     (iii) assuming  large  $N$  limit, i.e.  the corresponding   factorization  of the  correlators 
     $\bra J_s ... J_s \ket $    one can represent the original partition function in terms of 
     the path integral  over $\p_s$   with  the action $N \int  d^D x d^D x' \p_s(x)  K_s(x,x') \p_s(x') $,   
     where  $  K_s(x,x') = \bra J_s (x)  J_s (x') \ket$   or,  by conformal invariance,   
     $K_s (x,x')   \sim   P_s   |x-x'|^{-2 \Delta_-}$  (with $|x-x'|$    replaced by a geodesic  distance 
     in the case of  $S^D$); (iv) one observes that $K_s(x,x')$  is the Green's function for the  CHS operator 
     $\D_s = P_s  \del^{ 2 s + D-4}$,  i.e. $ \D_s K_s(x,x') = P_s \delta^{D}(x-x')$; \\
     (v) this  gives a  relation between ratio of  the  IR  and   UV  CFT  partition functions  and the CHS partition function,  
     $Z(\infty)/Z(0) = Z^{-1}_{\rm CHS} + O(1/N)$; (vi)   since AdS/CFT   should imply  the 
     equivalence of the CFT  partition function on $S^D$ and  the partition function of the 
     AdS  theory,  and  noting that the only  term that survives  in the ratio $Z(\infty)/Z(0) $ is 
     the spin $s$  field contribution  (as  this is the only field 
      that changes  its dimension and thus the corresponding  boundary conditions), 
     one is then led to the  relation \rf{zz} (taken in power -1).
     
     One  may  also  run  this argument in the opposite direction: starting with a  deformation of the  free IR CFT 
     by  the irrelevant  operator $\td \g \td J_s^2$   with $\td J_s$ having canonical dimension $\Delta_+= s + D-2 $. 
     In this case  we get the deformed CFT 
     partition function $\td Z(\td \g)$  expressed in terms of 
      the integral over the corresponding  auxiliary field $\td \p_s$  with the action 
      $N \int  d^D x d^D x' \td  \p_s(x) \td  K_s(x,x') \td \p_s(x') $, where 
     $\td  K_s (x,x') = \bra\td  J_s (x) \td  J_s (x') \ket    \sim   P_s   |x-x'|^{-2 \Delta_+}$. 
     Using  that the singular part of this  kernel  is  proportional to 
      the CHS operator  $\D_s = P_s  \del^{ 2 s + D-4}$, i.e. 
     $\td  K_s (x,x')\big|_{|x- x'| \to \eps}  \sim \ln \eps\  \D_s   \delta^{(D) } (x,x')  $ 
     we then end up with $\td Z(\infty ) /\td Z(0) =  Z_{\rm CHS} + O(1/N)$, i.e. 
     the same relation  as above since $\td Z(\infty ) /\td Z(0) = Z(0)/Z(\infty)$.
    Note that  this  consistency depends on a specific  regularization, allowing to ignore the finite part
    in $\td  K_s (x,x')$.

The equality    of the $\aa_s$   coefficients \rf{19}  governing the logarithmic  divergences  of these partition functions
  (found  in \ci{gio} as the IR   divergence of the   AdS$_{5}$  part of \rf{zz}  and  found here
  as the UV   divergence of  the  S$^4$ part of  \rf{zz})
  provides  a   first check of this equality,  and  it should  be possible also
  to show  the equality of the finite parts (using an appropriate regularization as in \ci{dorn,diaz,gio}). 


 \iffa
The  conformal higher spin  field  term  appears as ``induced'' ones   at the boundary \ci{lt,tse} -- as  coefficient of the logarithmic term
in the two-point function of the corresponding conserved  higher spin current,
$ \langle J_{m_1...m_s}  (q) J_{n_1...n_s} (-q) \rangle \sim  P_s(q)   q^{2s}  \log {q^2\ov L^2}$.
 \fi






\section*{Acknowledgments}
We are  grateful  to   S. Giombi and R. Metsaev  for important  discussions,  suggestions  and
initial collaboration.
We also thank   V. Didenko,   M. Grigoriev,  I. Klebanov,   G. Korchemsky, K. Mkrtchyan, R. Roiban, 
 E. Skvortsov   and M. Vasiliev  for useful discussions,  
and S. Giombi, I. Klebanov  and R. Metsaev  for   helpful  comments on the draft.
This  work was supported by the ERC Advanced grant No.290456
and also by the STFC grant
ST/J000353/1.

\np
 \appendix

\section*{Appendix A:   Notation and some useful relations }

\refstepcounter{section}
\def\theequation{A.\arabic{equation}}
\setcounter{equation}{0}

In this paper we always use symmetric traceless   tensors (or $\g$-traceless   spinor-tensors)
and so do not explicitly indicate the  tracelessness condition.

The curvature  tensor of  conformally-flat   Einstein  backgrounds  that we assume is
\be
R_{mnkl} = \ep ( g_{mk} g_{nl} - g_{mk} g_{nl} )  \   , \ \ \ \ \ \
\ep= \pm  r^{-2} = { 2 \L \ov (D-1) (D-2) }  \ ,  \la{a1} \ee
where   $\ep >0$  for $dS_{D}$ (or S$^D$ in the case of euclidean signature)
 and $\ep < 0$ for AdS$_D$ spaces.
In $D=4$  one has  $\ep= {\L \ov 3}, \ R= 4 \L = 12 \ep$.
One may  assume   that the curvature radius   is $r=1$ so that $\ep=\pm 1$.

Generic   covariant second-order  operator defined  on rank $s$ tensors in
such constant curvature space  can be put into the form
\be
\la{a2}
\dn_s =  \dn_s(M^2) \equiv   - \na^2_s + M^2\ep   \ .
\ee
 $  \dn_{s\pe}(M^2)$   will stand for $  - \na^2 + M^2\ep  $ defined   on   transverse traceless   tensors of  rank $s$.
In general (modulo zero-mode  contributions)
\be \la{dd}
\det \dn_{s}(M^2 ) =   \det \dn_{s\perp}(M^2)\, \det \dn_{s-1}  ( M^2-  (2s+ D-3 ) ) \ .
\ee
In particular, in $D=4$
\be &&\det \dn_{2}(M^2) =   \det \dn_{2\perp}(M^2)\, \det \dn_1 ( M^2- {\te{5}} ) \ , \\ &&  \det \dn_{1}(M^2 ) =   \det \dn_{1\perp}(M^2)\, \det \dn_0 ( M^2-  3) \ .  \la{ddd}\ee
 in agreement with \rf{ddd}.

One may  decompose
$ \p_{m_1...m_s} = \p_{m_1...m_s\perp}
+ [  \na_{(m_1} \xi_{m_2 ... m_s)} -  {\rm traces}] $
and compute the Jacobian of the corresponding transformation.
For example,
\be  && \p_m = \p_{m\pe} + \na_m\xi \ , \ \ \ \ \ \ \ \ \ \ J_1=[\det \De_0(0)]^{1/2}  \ ,  \\
 && \p_{mn} =\p_{mn\pe} + \na_{(m} \xi_{n) \pe} +   (\na_m\na_n - D^{-1} \na^2) \xi \ , \\
&&
 J_2= [\det \De_{1\pe} (    - D+1   ) \   \det \De_{0} (    -  D  )]^{1/2}\   [ \det \De_{0} (   0 )   ]^{1/2}  \ . \la{hj}
\ee
Here $ \De_{1} (    - D+1   )$ is the familiar ghost operator
 $\De_{1mn}=- \na^2_{mn} - R_{mn} $    in the case of \rf{a1}. 
 The first factor in \rf{hj}  is the
 ghost determinant, while  the second factor  is cancelled against similar   factor  in   the  volume of the gauge group
 (the gauge group vector  parameters are not transverse).

 Similar   decomposition for the  rank 3 tensor $\p_{mnk}=(  \p_{mnk\perp}, \xi_{mn\perp}, \xi_{m\perp},\xi)$   gives
\bea
&&J_3= [\det \dn_{2\pe} (    - 2D  ) \   \det \dn_{1\pe } (    -  2D-1  )\ \ \det \dn_0({-2D-2 })]^{1/2}\no  \\
&& \ \ \ \ \ \ \ \   \times \ [ \det \dn_{1\pe } (    -  D +1  )\ \det \dn_0{(- D  )} \ \det \dn_{0 }  (0)        ]^{1/2}  \ .  \la{a3}
\eea
Again, the first factor is  the ghost determinant  (for  details in the $D=3$ case see \ci{gab}).

Similar relations  are found for  the fermions,   e.g., for the $D=4$ gravitino 
\be \la{a9}
\psi_m= \psi_{m\pe} + (\na_m - \fo \g_m \g^k \na_k)  \xi \ , \ \ \ \ \ \ \ \ \ \
J_{\tri}  =  \det \dn_{\hav} (-1) \ .
\ee

\def \Del {\Delta}

\section*{Appendix B:  Comments on higher-order conformal  scalar operators}

\refstepcounter{section}
\def\theequation{B.\arabic{equation}}
\setcounter{equation}{0}

In  this paper we  considered  a family  of conformal higher-spin operators   defined  
on {\it symmetric traceless   tensors} $\p_s$, i.e.  
\be 
S=\int  d^D x \ \sqrt g \      \p^{m_1 ... m_s}   ( \nabla^{2s + D -4}  + ...)    \p_{m_1 ... m_s} \ , \la{b1} 
\ee
where $\p_s$ has dimension $ 2-s$.
 We suggested that  on Einstein-space background 
this  kinetic  operator   takes factorized form like \rf{pass}  in $D=4$.\foot{Let us note    
that in addition to $(2s + D-4)$-derivative  
CHS  operators discussed in this paper   one may formally  consider also  
operators  defined on symmetric traceless tensors with smaller number of derivatives
that have  Weyl symmetry  (see, e.g.,  
   \ci{dn2,rom,erd}  for some  2-derivative cases  and    \ci{sha,jm} for  general discussions). 
 Such 2-derivative  operators 
 effectively  appear  in the factorisation of  CHS operators   on Einstein background
 (as $k=0$   term in \rf{pass} in $D=4$ \ci{dn2,dw3}).
  In general $D$ the operator discussed in \ci{erd}  restricted to  transverse 
  fields and   considered on conformally-flat backgrounds
  corresponds to the ``maximal negative depth''  operator in \rf{pm},\rf{maa} with 
  $k=-\ha (D-4)$  or $i=1$ in \rf{exx},\rf{ex11}, i.e.  is  given by 
  $-\nabla^2_s + [ s +  {1\ov 4} D (D-2) ] \ep$. 
  }

Similar factorization is known  for higher-order conformal  operators    $\Del_{(2r)}$
defined on {\it scalars}
\be 
S_{2r} =\int  d^D x \ \sqrt g \      \p \,  \Del_{(2r)} \,  \p  \ , \ \ \ \ \ \ \ \ \ \ \ \ \ 
  \Del_{(2r)}  = (- \nabla^2)^{r}   + ...  \ . 
 \la{b111} 
\ee
In fact, a  special  case of such  scalar  action  with $r= \ha ( D-4)$   appears  
as  the special  case of  \rf{b1}    with $s=0$. 

In general, 
in  addition to the  familiar  conformal scalar operator \be \Del_{(2)}=- \na^2 + { D-2\ov 4(D-1)} R\ ,  \la{b0}\ee
one may define also $\Del_{(4)}= \na^4 +...$, etc.  In $D=4$  there  are two choices:
 $\Del_{(2)}=- \na^2 + { 1\ov 6} R $  and $\Del_{(4)}$   introduced  in \ci{ft1}
  and independently (for $D\geq 4$) in \ci{pan}:
\be 
S_4=\int  d^4 x \ \sqrt g \  \p\  \Del_{(4)}\ \p =      \int  d^4 x \ \sqrt g \     \Big[ \na^2  \p  \na^2 \p 
-   2 ( R^{mn} - {\te{1 \ov 3}} g^{mn} R) \na_m \p \na_n \p \Big]   \ . \la{b2} 
\ee
 As is  clear from \rf{b2}, on Ricci-flat   background $ \Del_{(4)}= ( \Del_{(2)})^2 =( \na^2)^2$
 while on the constant curvature   background \rf{a1}  one gets 
  $\Del_{(2)}=- \na^2 + { 1\ov 6} R  = -\na^2  + 2 \ep$   and 
 \be   \Del_{(4)}=  (-\na^2) ( -\na^2  + 2 \ep) =   \dn_0(0) \dn_0(2) \ , \la{bb2} \ee
 where we used the notation in \rf{a2}.
 This  leads to  a simple derivation of the corresponding Weyl anomaly  in \rf{1} \ci{ft1,ft2,ftr}:\foot{A discussion 
 of this anomaly in mathematics literature appeared also  in \ci{br1}.}
 it is just a sum of   anomalies  of  minimal  scalar Laplacian  $-\na^2$  and  the conformally-coupled one, i.e. 
 $\b_1= 2 \b_1 [-\na^2]=   { 1 \ov 90}, \   \ \aa= \aa[  \dn_0(0)] + \aa[ \dn_0(2)] =  - { 7 \ov 90} $ (and thus  $\cc= - { 1 \ov 15}$). 
 
  In  any  even   dimension $D$  one  may consider a family of 
higher-order conformal scalar  Laplacians       $\Del_{(2)}, \Del_{(4)}, \Del_{(6)}, ... , \Del_{(D)}$
 \ci{gjms}  usually referred  to as   GJMS   operators. 
 The operator $\Del_{({ D-4\ov 2})}$ appearing   in  the $s=0$ case of \rf{b1} is thus a special member of this family. 
 
 In the case of a  Ricci-flat   background  the   GJMS  operator $\Del_{(2r)}$  is simply the $r$-th  power of the   standard Laplacian $-\na^2$ 
 while for a   conformally-flat background (a   sphere $S^{D}$) the   operators   $\Del_{(2r)}$   were explicitly constructed   in \ci{br2}
 by arguing that they  should  take the following factorized  form\foot{Here  we set   $\ep=1$ in \rf{a1}    so that 
 $\Del_{(2)}=- \na^2 + { D-2\ov 4(D-1)} R  =  -\na^2 +  \fo D ( D -2) $. }
 \be 
&&  \Del_{(2r)} = \prod_{i=1}^{r}  \big[ \Del_{(2)} -  i(i-1)  \big] =
  \prod_{i=1}^{r}  \dn_0 (m^2_{{i}}) \ ,\qquad  \quad \dn_0 (m^2_{i})  =  - \na^2 +   m^2_{{i}}\ep\   , \la{b3}\\
 && \ \ \ \ \ \ \ \ 
 m^2_{i} = \fo D ( D-2)  - i(i-1)= ( \ha D  -  i ) (\ha D     + i-1)  \ .  \la{b4} \ee 
 The operator \rf{b3}   is  positive  for  $ r \leq \ha D$.
 Eq.\rf{b3} is thus   a   generalization of the relation  \rf{bb2}   in the   $r=2$    case. 

 The  factorized structure of \rf{b3}  is  obviously similar to that  in the case of higher-spin conformal operators in \rf{pass},\rf{maa}. 
 The direct derivation of \rf{b3}  in \ci{gr}
 that uses stereographic projection from flat space  should certainly 
    have  an analog in the  CHS case of \rf{pass}.

 The representation \rf{b3}   has a straightforward  generalization \ci{gov} 
 to any  Einstein-space background
 $R_{mn} = {1 \ov D} R g_{mn}$:
 \be 
&&  \Del_{(2r)} = \prod_{i=1}^{r}  \Big[   - \na^2 +  q_{{i}}
     R  \       \Big]  \  , \ \ \ \ \ \ \  \ \ \ \ \ \ \ 
      q_{{i}}=  { (\ha D     -i ) ( \ha D   + i-1  ) \ov  D (D-1)   }\ , \la{b5}
\ee
so  that, in particular, $ \Del_{(D)} = [- \na^2 + { D-2\ov 4(D-1)} R ] \ ... \  [ -\na^2] $.
For some other general properties of GJMS operators see also \ci{juhl}.\foot{See  also  \ci{dowk}   for a   computation of the corresponding conformal anomaly  and  determinant  on $S^D$.}

\bigskip

\end{document}


\bibitem{bast}
  F.~Bastianelli, R.~Bonezzi, O.~Corradini and E.~Latini,
  ``Effective action for higher spin fields on (A)dS backgrounds,''
  JHEP {\bf 1212}, 113 (2012)
  [arXiv:1210.4649].